\documentclass[twoside]{article}

\usepackage[accepted]{aistats2023}
\newcommand{\G}{{\cal G}}

\def\ci{\perp\!\!\!\perp}

\DeclareMathOperator{\pa}{pa}
\DeclareMathOperator{\nb}{nb}

\newtheorem{thm}{Theorem}
\newtheorem{prop}{Proposition}
\newtheorem{cor}{Corollary}
\newtheorem{lem}{Lemma}

\newenvironment{thma}[1]{\par\noindent{\bf Theorem #1\ }\em}{\em}

\newenvironment{proa}[1]{\par\noindent{\textbf{Proposition #1} }\em}{\em}

\newenvironment{prf}{\noindent\textit{Proof:}\begin{mdseries}}{\end{mdseries}{\hfill\scriptsize$\Box$}}

\usepackage[normalem]{ulem}
\newcommand{\stkout}[1]{\ifmmode\text{\sout{\ensuremath{#1}}}\else\sout{#1}\fi}

\usepackage[round]{natbib}

\usepackage{color}

\usepackage{amssymb,amsmath}

\usepackage{tikz,pgf}

\usetikzlibrary{shapes.arrows,shapes}

\begin{document}

%

%

\twocolumn[

\aistatstitle{The Lauritzen-Chen Likelihood For Graphical Models}

\aistatsauthor{ Ilya Shpitser }

\aistatsaddress{Department of Computer Science\\
  Johns Hopkins University\\
  Baltimore, MD, 21218\\
  \texttt{ilyas@cs.jhu.edu}\\
}

]

\begin{abstract}
Graphical models such as Markov random fields (MRFs) that are associated with undirected graphs, and Bayesian networks (BNs) that are associated with directed acyclic graphs, have proven to be a very popular approach for reasoning under uncertainty, prediction problems and causal inference.

Parametric MRF likelihoods are well-studied for Gaussian and categorical data.  However, in more complicated parametric and semi-parametric settings, likelihoods specified via clique potential functions are generally not known to be congenial {(jointly well-specified)} or non-redundant.  Congenial and non-redundant DAG likelihoods are far simpler to specify in both parametric and semi-parametric settings by modeling Markov factors in the DAG factorization.  However, DAG likelihoods specified in this way are not guaranteed to coincide in distinct DAGs within the same Markov equivalence class.  This complicates likelihoods based model selection procedures for DAGs by ``sneaking in'' potentially unwarranted assumptions about edge orientations.

In this paper we link a density function decomposition due to Chen with the clique factorization of MRFs described by Lauritzen to provide a general likelihood for MRF models.  The proposed likelihood is composed of variationally independent, and non-redundant closed form functionals of the observed data distribution, and is sufficiently general to apply to arbitrary parametric and semi-parametric models.  We use an extension of our developments to give a general likelihood for DAG models that is guaranteed to coincide for all members of a Markov equivalence class.  Our results have direct applications for model selection and semi-parametric inference.
\end{abstract}

\section{INTRODUCTION}
\label{sec:intro}

Graphical Markov models are a widely used approach for probabilistic modeling tasks of all types, including natural language and speech processing \citep{bilmes05graphical,smith11linguistic}, computational biology \citep{sachs05causal}, computer vision \citep{ji19probabilistic}, and causal inference \citep{spirtes01causation,pearl09causality}, among many others.  The popularity of graphical models stems from their tractable likelihoods expressed via factorizations, and intuitive graphical visualization of conditional independence restrictions in the model.
Common graphical models include Markov random fields (MRFs) \citep{ising25beitrag,sherrington75solvable}, associated with undirected graphs (UGs), and Bayesian networks (BNs) \citep{pearl88probabilistic}, associated with directed acyclic graphs (DAGs).

MRF likelihoods for Gaussian and categorical data are very well studied \citep{lauritzen96graphical}.  In modern high dimensional applications \citep{handbook19graphical}, MRF likelihoods are specified by modeling terms in the clique factorization of the model: $p(\vec{v}) = \prod_{\vec{C}} \phi_{\vec{C}}(\vec{v}_{\vec{C}})$, where the product is over potential functions corresponding to cliques in an undirected graph, and $\vec{v}_{\vec{C}}$ is the value assignment $\vec{v}$ restricted to $\vec{C} \subseteq \vec{V}$.  Terms $\phi_{\vec{C}}$ in such likelihoods are not, in general, known to be congenial and non-redundant.

Congeniality refers to the property that different parts of the model are not specified in a mutually contradictory way.  Congeniality was first described in the context of multiple imputation in missing data problems, where careless use of analysis and missing data models may impose contradictory restrictions, guaranteeing model misspecification \citep{meng94multiple}.  However, lack of congeniality becomes a concern whenever different parts of the model specification ``overlap'' and impose different restrictions on the same set of variables, as may happen with clique factors of an MRF.

A likelihood is non-redundant if different parts of the likelihood specification do not carry the same information about the model.  Redundant likelihoods may be undesirable as they do not have identified parameterizations, and do not yield insight about the model dimension or the model tangent space.  Incongenial and redundant likelihoods are described in more detail below.

Though in many applications congenial and non-redundant likelihoods are not needed, they are crucial if model parameters are of primary interest.  In addition, such likelihoods are important for deriving semi-parametrically efficient {estimators based on} influence functions that take advantage of Markov restrictions in 
{graphical models} \citep{tsiatis06missing}, and for score based model selection algorithms \citep{chickering02equiv}.

In this paper, we describe general likelihoods for MRFs based on terms of the clique factorization that are guaranteed to be congenial and non-redundant.  These terms are defined algebraic functions of the observed data distributions termed \emph{partial cross-product ratios} by \cite{lauritzen96graphical} that may be viewed as generalizations of odds ratio functions, or as generalized higher order interaction functions.  The likelihood is based on a synthesis of ideas on clique factorizations presented by 
\cite{lauritzen96graphical}, and the odds ratio decomposition derived by 
\cite{chen07semiparametric}.
We use this likelihood to define parametric and semi-parametric likelihoods for BN models that are guaranteed to be identical within distinct DAGs that are Markov equivalent (obey the same set of marginal and conditional independence restrictions).

After introducing some preliminaries {in Section~\ref{sec:prelims} and illustrating the concepts of congeniality, non-redundance, and variational independence in Section~\ref{sec:terms},} we describe the Chen and Lauritzen decompositions of conditional distributions in Section~\ref{sec:decomp}, describe the main results showing the close relationship of these decompositions in Section~\ref{sec:connections}, and describe BN likelihoods in Section~\ref{sec:dags}.  We conclude with some examples of the derived likelihoods, and discuss limitations and areas of future work.

\section{PRELIMINARIES}
\label{sec:prelims}



We first introduce necessary graphical modeling preliminaries.  
Graphs are assumed to have a vertex set $\vec{V}$, and we will restrict attention to positive distributions. 
Given any graph $\G$, for $\vec{S} \subseteq \vec{V}$, an induced subgraph $\G_{\vec{S}}$ of $\G$ is defined as the graph with a vertex set $\vec{S}$ and all edges in $\G$ connecting elements in $\vec{S}$.

Given an undirected graph (UG) $\G$, a clique $\vec{C}$ is a (possibly empty) subset of vertices in $\vec{V}$ that are pairwise connected in $\G$.  The set of all cliques in $\G$ is denoted by ${\cal C}(\G)$, while the set of all \emph{maximal} cliques is denoted by $\bar{\cal C}(\G)$.  Note that,  in general, neither ${\cal C}(\G)$ nor $\bar{\cal C}(\G)$ will form a partition of $\vec{V}$ in $\G$.  A joint distribution $p(\vec{v})$ is in the Markov random field (MRF) model of a UG $\G$ if for every value $\vec{v}$, $p(\vec{v}) = \prod_{\vec{C} \in {\cal C}(\G)} \phi_{\vec{C}}(\vec{v}_{\vec{C}})$, where 
$\phi_{\vec{C}}$ are \emph{potential functions} which map values of $\vec{C}$ to real numbers.  Potential functions are \emph{not} necessarily normalized probabilities.  Equivalently, $p(\vec{v})$ is in the MRF model if
$p(\vec{v}) = Z^{-1} \prod_{\vec{C} \in \bar{\cal C}(\G)} \phi_{\vec{C}}(\vec{v}_{\vec{C}})$, where $Z$ is a normalizing constant.

If we restrict attention to positive distributions, an MRF model may be equivalently defined as the set of distributions $p(\vec{v})$ that satisfy either the global or pairwise Markov property for $\G$.
The global Markov property for $p(\vec{v})$ and a UG $\G$ states that for any disjoint subsets $\vec{A}$,$\vec{B}$,$\vec{C}$ of $\vec{V}$ whenever all paths from $\vec{A}$ to $\vec{B}$ in $\G$ are intercepted by $\vec{C}$, then $\vec{A} \ci \vec{B} | \vec{C}$ in $p(\vec{v})$.  The pairwise Markov property for $p(\vec{v})$ and $\G$ states that for any vertex pair $A,B$ non-adjacent in $\G$, $A \ci B | \vec{V} \setminus \{ A, B \}$ in $p(\vec{v})$.

A joint distribution $p(\vec{v})$ is in the Bayesian network (BN) model of a directed acyclic graph (DAG) $\G$ if for every value $\vec{v}$, $p(\vec{v}) = \prod_{V \in \vec{V}} p(\vec{v}_{\{V\}} | \vec{v}_{\pa_{\G}(V)})$, where
$\pa_{\G}(V)$ are the set of parents of $V$ in $\G$.
A BN model may equivalently defined as the set of distributions $p(\vec{v})$ that satisfy the global Markov property for $\G$ given by the d-separation criterion \citep{pearl88probabilistic}, where
for any disjoint subsets $\vec{A}$,$\vec{B}$,$\vec{C}$ of $\vec{V}$ whenever all paths from $\vec{A}$ to $\vec{B}$ in $\G$ are d-separated by $\vec{C}$, then $\vec{A} \ci \vec{B} | \vec{C}$ in $p(\vec{v})$.


Both MRF and BN models may be generalized into \emph{conditional MRF (CMRF)} \citep{lafferty01conditional} and \emph{conditional BN (CBN)} models, associated with conditional undirected graphs (CUGs) and conditional directed acyclic graphs (CDAGs).  A CUG $\G(\vec{V},\vec{W})$ is a mixed graph containing \emph{random vertices} $\vec{V}$ and \emph{fixed vertices} $\vec{W}$ as well as undirected and directed edges, such that undirected edges are only among elements of $\vec{V}$ and directed edges are always from an element in $\vec{W}$ to an element in $\vec{V}$.  Similarly, a CDAG $\G(\vec{V},\vec{W})$ is a directed graph containing \emph{random vertices} $\vec{V}$ and \emph{fixed vertices} $\vec{W}$, such that there are no directed cycles, and the only allowed edges adjacent to $\vec{W}$ are out of elements in $\vec{W}$ and into elements in $\vec{V}$.  The notion of parents generalizes to conditional graphs to potentially include fixed vertices in $\vec{W}$.

A distribution $p(\vec{v} | \vec{w})$ is in the CMRF model of a CUG $\G(\vec{V},\vec{W})$ if for every $\vec{v},\vec{w}$,
$p(\vec{v}|\vec{w}) = \prod_{\vec{C} \in {\cal C}(\G_{\vec{V}})} \phi_{\vec{C}}(\vec{v}_{\vec{C}},\vec{w}_{\vec{C}^*})$, where for every element $\vec{C}$, $\vec{C}^* = \bigcap_{C \in \vec{C}} \pa_{\G}(C)$.
Equivalently, $p(\vec{v} | \vec{w})$ is in the CMRF model of a CUG $\G(\vec{V},\vec{W})$ if for every $\vec{v},\vec{w}$,
$p(\vec{v}|\vec{w}) = Z(\vec{w})^{-1} \prod_{\vec{C} \in \bar{\cal C}(\G_{\vec{V}})} \phi_{\vec{C}}(\vec{v}_{\vec{C}},\vec{w}_{\vec{C}^*})$, where $Z(\vec{w})$ is a \emph{normalizing function}.
A distribution $p(\vec{v} | \vec{w})$ is in the CBN model of a CDAG $\G(\vec{V},\vec{W})$ if for every $\vec{v},\vec{w}$,
$p(\vec{v}|\vec{w}) = \prod_{V \in \vec{V}} p(\vec{v}_{\{V\}} | \vec{v}_{\pa_{\G}(V) \cap \vec{V}}, \vec{w}_{\pa_{\G}(V) \cap \vec{W}})$.

\section{CONGENIALITY, NON-REDUNDANCE AND VARIATIONAL INDEPENDENCE}
\label{sec:terms}

Consider a non-parametric (unrestricted) model on two binary random variables, $A$ and $B$.
A variationally independent parameterization for the likelihood of this model may be obtained by choosing any three joint distribution parameters, for example $p(A=0,B=0)$, $p(A=0,B=1)$, and $p(B=1,A=0)$.
Variational independence means that any of these three parameters may be chosen to be any valid probability without restricting values of the other parameters.

An alternative parameterization for this likelihood is given by the following three parameters: $p(A=0)$, $p(B=0)$ and $p(A=0,B=0)$.  This \emph{M\"obius parameterization} is variationally dependent, since allowed values of the marginal parameters $p(A=0)$ and $p(B=0)$ may be influenced by the value of the joint parameter $p(A=0,B=0)$.  Despite being variationally dependent, this is a valid parameterization for categorical data and corresponds to a smooth parameter map given by the M\"obius inversion formula.  Variationally dependent parameterizations may require extensions to standard iterative optimization methods since any local step that changes parameter values (for instance via gradient descent) has to also ensure the newly chosen parameter values correspond to an element in the model.  M\"obius parameterizations have previously been provided for mixed graph models of categorical data \citep{evans14markovian,evans18smooth}.

Now consider a model on two real-valued random variables $A$ and $B$.  A generalization of the M\"obius parameterization would aim to specify certain parts $g_A(A)$, $g_B(B)$ and $g_{A,B}(A,B)$ of density functions $f(A)$, $f(B)$ and $f(A,B)$, respectively.  An \emph{incongenial} specification of objects $g_A,g_B$, and $g_{A,B}$ is one that imposes contradictory restrictions such that the resulting model (set of distributions) excludes the true distribution by construction.  In other words, an incongenial specification results in a misspecified model.  An extreme case of incongeniality is one where there is no valid density that satisfies all imposed restrictions, leading to the degenerate empty set model.  Issues of congeniality were first described when discussing multiple imputation approaches to missing data problems \citep{meng94multiple}.

A redundant specification of the model is one that uses ``too many pieces,'' leading to non-identification.  For example, a non-parametric model on two binary random variables $A$ and $B$ has dimension $3$.
A parameterization of this model that involves $4$ unrestricted real-valued parameters leads to non-identification: there are infinitely many values of these $4$ parameters that are consistent with any given distribution
$p(A,B)$.  The same issue may arise in semi-parametric models (despite such models being infinite dimensional).  A redundant specification of such a model leads to a situation where many specifications are consistent with a particular observed data distribution in the model.  In other words, a redundant model specification leads to a parameter map that is not bijective, which prevents identification of model specifications from the observed data distribution.  While redundant specifications may be harmless in some applications, they present difficulties in settings where model parameters (and thus their identification) is of primary interest.

In the remainder of this paper, we will describe an approach that leads to congenial, non-redundant and variationally independent specifications of a wide class of graphical models, including those associated with undirected, directed, and mixed graphs.

\section{LIKELIHOOD DECOMPOSITIONS}
\label{sec:decomp}

\cite{chen07semiparametric} has considered the following decomposition for an arbitrary conditional distribution $p(v_1, v_2 | w)$, given an arbitrary set of reference values $v_1^*, v_2^*$:
\begin{align}
p(v_1,\! v_2 | \vec{w}) \!=\!
\frac{
p(v_1 | v_2^*, \vec{w}) p(v_2 | v_1^*, \vec{w}) OR(v_1, v_2; v_1^*, v_2^* | \vec{w})
}{
Z(\vec{w})
},
\label{eqn:or-2}
\end{align}
where $OR(v_1, v_2; v_1^*, v_2^* | \vec{w}) = \frac{ p(v_1, v_2 | w) p(v_1^*, v_2^* | \vec{w}) }{ p(v_1, v_2^* | \vec{w}) p(v_1^*, v_2 | \vec{w}) }$ is the conditional odds ratio function.  In subsequent discussion, we will suppress the reference values from odds ratio functions, and denote them as e.g., $OR(v_1, v_2 | \vec{w})$.

The following important result is derived by \cite{chen03note,chen07semiparametric}.
In the interests of being self-contained, we reproduce the proof of this result in the Appendix.
\begin{prop}
The terms in the numerator of (\ref{eqn:or-2}) are \emph{non-redundant} and \emph{variationally independent}.
\label{prop:chen}
\end{prop}

The decomposition in (\ref{eqn:or-2}) can be generalized in a straightforward way to a multivariate conditional distribution $p(\vec{v} | \vec{w}) = p(v_1, \ldots, v_K | \vec{w})$ by inductively applying (\ref{eqn:or-2}) while taking $v_k$ as the first variable and $(v_{1}, \ldots, v_{k-1})$ as the second variable for all $k = 2, \ldots, K$.  This yields the following decomposition for $p(v_1, \ldots, v_K | \vec{w})$:
\begin{align}
\notag
Z(\vec{w})^{-1}
\cdot
\left(
\prod_{k=1}^K
p(v_k | \vec{v}_{-k}^*, \vec{w})
\right) \times \\
\times
\left(
\prod_{k=2}^K OR(v_k, (v_1, \ldots, v_{k-1}) | v_{k+1}^*, \ldots, v_K^*,\vec{w})
\right),
\label{eqn:or-k}
\end{align}
where $\vec{v}_{-k} \equiv (v_1, \ldots, v_{k-1}, v_{k+1}, \ldots, v_K)$.

Two properties of the decomposition in (\ref{eqn:or-k}) are worth noting.  First, by inductively applying Proposition \ref{prop:chen}, it is straightforward to establish that all terms in the numerator of (\ref{eqn:or-k}) are non-redundant, and variationally independent.  Second, this decomposition is valid for any order on variables in $\vec{V}$.

We now consider an alternative decomposition of a conditional distribution $p(\vec{v} | \vec{w})$.  The presentation is a generalization of a decomposition of joint distributions $p(\vec{v})$ described by \cite{lauritzen96graphical}.
For any subset $\vec{C} \subseteq \vec{V}$, define
\begin{align}
\notag
H_{\vec{C}}(\vec{v}_{\vec{C}}, \vec{w}) &\equiv \log p(\vec{v}_{\vec{C}}, \vec{v}^*_{\vec{V} \setminus \vec{C}} | \vec{w});\\
\hspace{0.5cm}
\tilde{\phi}_{\vec{C}}(\vec{v}_{\vec{C}}, \vec{w}) &\equiv \sum_{\vec{B} \subseteq \vec{C}} (-1)^{|\vec{C} \setminus \vec{B}|} H_{\vec{B}}(\vec{v}_{\vec{B}}, \vec{w}).
\label{eqn:moebius-terms}
\end{align}
Since $H_{\vec{C}}$ and $\tilde{\phi}_{\vec{C}}$ are defined for every subset $\vec{C}$ of $\vec{V}$, they are related by the M\"obius inversion formula, as follows:
\begin{align}
\log p(\vec{v} | \vec{w}) = H_{\vec{V}}(\vec{v}, \vec{w}) = \sum_{\vec{C} \subseteq \vec{V}} \tilde{\phi}_{\vec{C}}(\vec{v}_{\vec{C}}, \vec{w}).
\label{eqn:moebius-inversion-log}
\end{align}
For details, consult Appendix A.3 of \cite{lauritzen96graphical}.

We can rewrite (\ref{eqn:moebius-inversion-log}) as:
\begin{align}
p(\vec{v} | \vec{w}) = \prod_{\vec{C} \subseteq \vec{V}} \exp \left\{ \tilde{\phi}_{\vec{C}}(\vec{v}_{\vec{C}}, \vec{w}) \right\} = \prod_{\vec{C} \subseteq \vec{V}} \phi_{\vec{C}}(\vec{v}_{\vec{C}}, \vec{w}),
\label{eqn:moebius-inversion}
\end{align}
where we define, for every $\vec{C}$:
\begin{align}
\notag
\phi_{\vec{C}}(\vec{v}_{\vec{C}}, \vec{w}) &\equiv \exp \left\{ \tilde{\phi}_{\vec{C}}(\vec{v}_{\vec{C}}, \vec{w}) \right\}\\
&=
\exp \left\{ \sum_{\vec{B} \subseteq \vec{C}} (-1)^{|\vec{C} \setminus \vec{B}|} H_{\vec{B}}(\vec{v}_{\vec{B}}, \vec{w}) \right\}.
\label{eqn:phi}
\end{align}
We term (\ref{eqn:moebius-inversion}) the Lauritzen decomposition of $p(\vec{v}|\vec{w})$.  The CMRF factorization with respect to a CUG $\G(\vec{V},\vec{W})$ is the Lauritzen decomposition where terms that do not correspond to cliques in $\G_{\vec{V}}$ disappear. 
It is simple to show that the CMRF factorization implies the CMRF version of the pairwise Markov property.  In fact, for positive distributions the converse is also true, due to the following. 

\begin{thm}[Hammersly-Clifford for conditional MRFs]
Assume a positive $p(\vec{v} | \vec{w})$ obeys the pairwise Markov property for a CUG $\G(\vec{V},\vec{W})$.
That is, for every $V \in \vec{V}$, $Z \in \vec{V} \cup \vec{W}$ such that $Z$ is non-adjacent to $V$ in $\G(\vec{V},\vec{W})$,
$p(v | (\vec{v} \cup \vec{w}) \setminus \{ v \})$ is only a function of $(\vec{v} \cup \vec{w}) \setminus \{ z \}$.
Then $p(\vec{v} | \vec{w})$ factorizes with respect to $\G(\vec{V},\vec{W})$.
That is,
\begin{align}
p(\vec{v} | \vec{w})
=
\prod_{\vec{C} \subseteq \bar{\cal C}(\G_{\vec{V}})} \phi_{\vec{C}}(\vec{v}_{\vec{C}}, \vec{w}_{\vec{C}^*}),
\label{eqn:cmrf-f}
\end{align}
where for every $\vec{C}$, $\vec{C}^* = \bigcap_{C \in \vec{C}} \pa_{\G}(C)$, and $\phi_{\vec{C}}$ are terms defined as in (\ref{eqn:moebius-inversion}).
\label{thm:h-c-cmrf}
\end{thm}
The proof, which is a simple extension of the proof of the MRF version of the Theorem derived by \cite{lauritzen96graphical}, is in the Appendix.

The two decompositions of a conditional distribution $p(\vec{v} | \vec{w})$ in the CMRF model of a CUG $\G(\vec{V},\vec{W})$ have desirable and complementary properties.
The Chen decomposition in (\ref{eqn:or-k}) contains non-redundant and variationally independent terms, whereas the Lauritzen decomposition in (\ref{eqn:cmrf-f})
only has non-trivial terms corresponding to cliques of an induced subgraph $\G_{\vec{V}}$, thus allowing it to represent CMRF Markov restrictions.
As we show in the next section, these two decompositions are related, allowing us to define general likelihoods for CMRF models by taking advantage of both properties at once.

\section{CONNECTIONS BETWEEN CHEN AND LAURITZEN DECOMPOSITIONS}
\label{sec:connections}

To illustrate the connection between the two decompositions described in the previous section, we compare the decompositions for a three-variable joint distribution
$p(v_1, v_2, v_3)$.
The Chen decomposition (\ref{eqn:or-k}) is:
\begin{align}
\notag
p(v_1, v_2, v_3)
&=
{\color{red}Z^{-1}}
p(v_1 | v_2^*,v_3^*) p(v_2 | v_1^*,v_3^*) p(v_3 | v_1^*, v_2^*) \\
& \hspace{0.5cm}
OR(v_1, v_2 | v_3^*)
OR(v_3, (v_1,v_2))
\label{eqn:chen-3}
\end{align}
The Lauritzen decomposition (\ref{eqn:moebius-inversion}) of $p(v_1, v_2, v_3)$ is:
\begin{align}
\notag
&
	\underbrace{
	\overbrace{{\color{red}p(v_1^*,v_2^*,v_3^*)}}^{\phi_{\emptyset}}
	\overbrace{
	\frac{
 		p(v_1 | v_2^*,v_3^*)
 	}{
		{\color{red}p(v_1^* | v_2^*,v_3^*)}
	}
	}^{\phi_{\{V_1\}}}
	\overbrace{
	\frac{
 		p(v_2 | v_1^*,v_3^*)
 	}{
		{\color{red}p(v_2^* | v_1^*,v_3^*)}
	}
	}^{\phi_{\{V_2\}}}
	\overbrace{
	\frac{
 		p(v_3 | v_2^*,v_1^*)
 	}{
		{\color{red}p(v_3^* | v_2^*,v_1^*)}
	}
	}^{\phi_{\{V_3\}}}
	}_{\text{red terms are equal to $Z^{-1}$ in (\ref{eqn:chen-3})}} \times
	\\
&
	\overbrace{
	OR(v_1,\! v_2 | v_3^*)
	}^{\phi_{\{V_1, V_2\}}}
	\underbrace{
	\overbrace{
	OR(v_1,\! v_3 | v_2^*)
	}^{\phi_{\{V_1,V_3\}}}
	\overbrace{
	OR(v_2,\! v_3 | v_1^*)
	}^{\phi_{\{V_2,V_3\}}}
	\overbrace{
	\frac{
		OR(v_1,\! v_2 | v_3)
	}{
		OR(v_1,\! v_2 | v_3^*)
	}}^{\phi_{\{V_1,V_2,V_3\}}}
	}_{OR(v_3, (v_1, v_2))}
\label{eqn:lauritzen-3}
\end{align}
The decompositions (\ref{eqn:chen-3}) and (\ref{eqn:lauritzen-3}) are equivalent since $OR(v_1, v_3 | v_2^*)
OR(v_2, v_3 | v_1^*)
\frac{
OR(v_1, v_2 | v_3)
}{
OR(v_1, v_2 | v_3^*)
} = OR(v_3, (v_1,v_2))$ (see the Appendix), and consequently $\frac{p(v_1^* | v_2^*, v_3^*) p(v_2^* | v_1^*, v_3^*) p(v_3^* | v_1^*, v_2^*)}{p(v_1^*, v_2^*, v_3^*)} = Z$.

To see that the univariate conditional terms, such as $p(v_1 | v_2^*, v_3^*)$, terms corresponding to two variable subsets of $\{ V_1, V_2, V_3 \}$, such as $\phi_{\{V_1, V_2\}}$, and the
term corresponding to $\{ V_1, V_2, V_3 \}$, namely $\phi_{\{V_1,V_2,V_3\}}$ in (\ref{eqn:lauritzen-3}) are non-redundant and variationally independent, we argue as follows. Note that the
terms in the numerator of the decomposition (\ref{eqn:chen-3}) are non-redundant and variationally independent for any order $(i,j,k)$ on indices of variables $V_1, V_2, V_3$.
This implies the univariate conditional terms, any term of the form $OR(v_i,v_j | v_k^*)$, and the other OR terms grouped with $OR(v_k, (v_i,v_j))$ are non-redundant and variationally independent.
This also implies these terms are non-redundant and variationally independent of the ``remainder'' of $OR(v_k, (v_i,v_j))$ (under any index order $i,j,k$) once two variable OR terms are excluded,
which is precisely $\phi_{\{V_1,V_2,V_3\}}$.

As the following results show, the relationship between the Chen and Lauritzen decompositions highlighted by this example is general.

\begin{thm}
Given a positive distribution $p(\vec{v} | \vec{w})$ on an ordered set of variables $\vec{V} \equiv \{ V_1, \ldots, V_{K} \}$,
for each $k = 2, \ldots, K$,
\begin{align}
\notag
OR(v_k, (v_1, \ldots, v_{k-1}) | v_{k+1}^*, \ldots, v_{K}^*, \vec{w})\\
= \prod_{\{ V_k \} \subset \vec{C} \subseteq \vec{V}_{{k}}} \phi_{\vec{C}}(\vec{v}_{\vec{C}},\vec{w}),
\label{eqn:chen-lauritzen}
\end{align}
where $\vec{V}_k \equiv \{ V_1, \ldots, V_k \}$, and $\phi_{\vec{C}}(\vec{v}_{\vec{C}},\vec{w})$ is defined as in (\ref{eqn:phi}).
\label{thm:chen-lauritzen}
\end{thm}
\begin{prf}
For any value vector $\{ v_1, \ldots, v_K \}$, denote $\vec{v}_{k} \equiv \{ v_1, \ldots, v_k \}$.
Denote
$OR(\vec{v}_{k-1},v_k | v_{k+1}^*, \ldots, v_{K}^*, \vec{w})$
by $\eta_k$.  By definition,
\begin{align*}
\eta_k
\!=\!
\frac{
p(\vec{v}_{k-1},\!v_{k},\! v_{k+1}^*, ..., v_K^* | \vec{w})
p(\vec{v}_{k-1}^*,\!v_{k}^*, \!v_{k+1}^*, ..., v_K^* | \vec{w})
}{
p(\vec{v}_{k-1}, \!v_{k}^*,\! v_{k+1}^*, ..., v_K^* | \vec{w})
p(\vec{v}_{k-1}^*,\! v_{k},\! v_{k+1}^*, ..., v_K^* | \vec{w})
}
\end{align*}
We claim that the following set of equalities hold.
\begin{align}
\notag
& \prod_{\{ V_k \} \subset \vec{C} \subseteq \{ V_1, \ldots, V_k \}}
\phi_{\vec{C}}(\vec{v}_{\vec{C}},\vec{w})\\
&=
\prod_{\{ V_k \} \subset \vec{C} \subseteq \{ V_1, \ldots, V_k \}}
\exp \left\{
\sum_{\vec{B} \subseteq \vec{C}} (-1)^{|\vec{C} \setminus \vec{B}|} H_{\vec{B}}(\vec{v}_{\vec{B}},\vec{w})
\right\}
\label{eqn:eta-phi-deriv}
\\
\notag
&=
\exp \left\{
H_{\{ V_1, \ldots, V_k \}} \!+\! H_{\emptyset}\! -\! H_{\{ V_k \}}\! -\! H_{\{ V_1, \ldots, V_{k-1} \}}
\right\} = \eta_k,
\end{align}
where $H_{\vec{C}}(\vec{v}_{\vec{C}},\vec{w})$ for any $\vec{C} \subseteq \vec{V}$ is defined as in (\ref{eqn:moebius-terms}).

The first equality holds by definition of $\phi_{\vec{C}}(\vec{v}_{\vec{C}},\vec{w})$.
To see that the last two equalities hold, we argue as follows.
First, the number of subsets of $\{ V_1, \ldots, V_k \}$ is $2^k$, and the number
of subsets of $\{ V_1, \ldots, V_k \}$ that includes $V_k$ is $2^{k-1}$.
Thus the number of terms in the first line of (\ref{eqn:eta-phi-deriv}) is $2^{k-1} - 1$ (since the set $\{ V_k \}$ does not have a corresponding term).

$H_{\{ V_1, \ldots, V_k \}}$ occurs once, in the term corresponding to $\{ V_1, \ldots, V_k \}$, with a positive sign.

Since all terms correspond to sets that must contain $V_k$, $H_{\{ V_1, \ldots, V_{k-1} \}}$ occurs once,
in the term corresponding to $\{ V_1, \ldots, V_k \}$, with a negative sign, since
$|\{ V_1, \ldots, V_{k} \} \setminus \{ V_1, \ldots, V_{k-1} \}| = 1$.

$H_{\emptyset}$ and $H_{\{ V_k \}}$ occur in every term, and moreover $2^{k-1}-2$ occurrences of these terms can
be paired with opposite signs, and thus cancel.  The last occurrence is in the term
corresponding to $\{ V_1, \ldots, V_k \}$, where $H_{\emptyset}$ occurs with a positive sign, and $H_{\{ V_k \}}$ occurs
with a negative sign.

Consider a set $\vec{B} \subseteq \{ V_1, \ldots, V_k \}$ that is not equal to $\emptyset$, $\{ V_k \}$, $\{ V_1, \ldots, V_{k-1} \}$, or $\{ V_1, \ldots, V_k \}$.
The sets $\vec{C}$ such that $\{ V_k \} \subset \vec{C} \subseteq \{ V_1, \ldots, V_k \}$ where $\vec{B} \subseteq \vec{C}$ is
$\vec{C}' \cup \vec{B}$, where $\vec{C}'$ is any subset of $\{ V_1, \ldots, V_k \} \setminus \vec{B}$.  It's clear that there is an even number of such
subsets, specifically $2^{| \{ V_1, \ldots, V_k \} \setminus \vec{B} |}$, and that the corresponding terms $H_{\vec{B}}$ cancel as they occur with
alternating signs.

The last equality then holds by the definition of $\eta_k$.
\end{prf}

\begin{cor}[Lauritzen-Chen {(L-C)} decomposition]
Any positive distribution $p(v_1, \ldots, v_K | \vec{w})$ can be written as,
\begin{align}
\label{eqn:chen-f}
\notag
&\hspace{0.5cm} 
Z(\vec{w})^{-1}
\left( \prod_{k=1}^K p(v_k | \vec{v}_{-k}^*,\vec{w}) \right) \times \\
&\hspace{0.5cm} \prod_{k=2}^K OR(v_k, (v_1, \ldots, v_{k-1}) | v_{k+1}^*, \ldots, v_K^*, \vec{w})
\end{align}
\begin{align}
\notag
&= \prod_{\vec{C} \subseteq \{ V_1, \ldots, V_K \}} \phi_{\vec{C}}(\vec{v}_{\vec{C}},\vec{w}) \times\\
\notag
&= \left( \frac{\phi_{\emptyset}(\vec{w})}{\prod_{k=1}^K p(v_k^* | \vec{v}_{-k}^*, \vec{w}) } \right) \left( \prod_{k=1}^K p(v_k | \vec{v}_{-k}^*, \vec{w}) \right)\\
&\hspace{0.5cm} \prod_{\vec{C} \subseteq \{ V_1, \ldots, V_K \}; |\vec{C}|\geq 2} \phi_{\vec{C}}(\vec{v}_{\vec{C}},\vec{w}).
\label{eqn:lauritzen-f}
\end{align}
Furthermore, $Z(\vec{w}) = \frac{\prod_{k=1}^K p(v_k^* | \vec{v}_{-k}^*, \vec{w}) }{\phi_{\emptyset}(\vec{w})}$ for all $\vec{w}$, and all terms
$p(v_k | \vec{v}_{-k}^*,\vec{w})$ (for all $k$) and $\phi_{\vec{C}}(\vec{v}_{\vec{C}},\vec{w})$ if $|\vec{C}| \geq 2$ in (\ref{eqn:lauritzen-f}) are non-redundant and
variationally independent.
\label{cor:chen-lauritzen}
\end{cor}
\begin{prf}
Equality of (\ref{eqn:chen-f}) and (\ref{eqn:lauritzen-f}), as well as equality of $Z(\vec{w})$ and $\frac{\prod_{k=1}^K p(v_k^* | \vec{v}_{-k}^*, \vec{w}) }{\phi_{\emptyset}(\vec{w})}$
follows by Theorem \ref{thm:chen-lauritzen}.

To show non-redundance and variational independence, we proceed by induction.
Results of \cite{chen03note,chen07semiparametric} show that all terms in the numerator of the right hand side of (\ref{eqn:chen-f}) are non-redundant and variationally independent.
Since $OR(v_1, v_2 | v_{3}^*, \ldots, v_{k}^*,\vec{w}) = \phi_{\{ V_1, V_2 \}}(\vec{v}_{\{V_1,V_2\}},\vec{w})$, and since the (\ref{eqn:chen-f}) holds for any ordering
on variables $\vec{V}$, we conclude that terms $p(v_k | \vec{v}_{-k}^*, \vec{w})$, $\phi_{\{ V_i, V_j \}}(\vec{v}_{\{ V_i, V_j \}},\vec{w})$ are non-redundant and variationally independent for any $i,j,k$.
This establishes the base case.

Assume we have shown that terms $p(v_k | \vec{v}_{-k}^*, \vec{w})$ (for any $k \in \{ 1, \ldots, K \}$) and $\phi_{\vec{C}}(\vec{v}_{\vec{C}},\vec{w})$ (for all $\vec{C}$ such that $|\vec{C}| < i$) are non-redundant and variationally independent.

Let $\eta_i \equiv OR(v_i, (v_1, \ldots, v_{i-1}) | v_{i+1}^*, \ldots, v_K^*, \vec{w})$.  Under the given variable ordering,
$\eta_{i}$ is non-redundant and variationally independent of all other terms in (\ref{eqn:chen-f}).  By Theorem \ref{thm:chen-lauritzen}, $\eta_{i}$ contains
a single term $\phi_{\vec{C}}(\vec{v}_{\vec{C}},\vec{w})$ of size $i$, and other terms of size smaller than $i$.  These terms, along with univariate conditionals in (\ref{eqn:chen-f}),
are all non-redundant and variationally independent of $\phi_{\vec{C}}(\vec{v}_{\vec{C}},\vec{w})$ by the inductive hypothesis.  Furthermore, $\phi_{\vec{C}}(\vec{v}_{\vec{C}},\vec{w})$ is non-redundant and variationally
independent of $\eta_{j}$ for all $j > i$.  Since variation independence in (\ref{eqn:chen-f}) is order-independent, the same argument
can be repeated for any set $\vec{C}$ of size $i$, and the corresponding term $\phi_{\vec{C}}(\vec{v}_{\vec{C}},\vec{w})$.  This establishes the induction, and thus the result.
\end{prf}

The Lauritzen-Chen decomposition immediately yields a non-redundant and variationally independent factorization of CMRFs, as follows.

\begin{cor}
{\bf (Lauritzen-Chen (L-C) factorization of conditional MRFs)}
For any positive $p(\vec{v}|\vec{w})$ in the CMRF model of a CUG $\G(\vec{V},\vec{W})$,
\begin{align}
\notag
p(\vec{v}|\vec{w}) &= 
\left( \frac{\phi_{\emptyset}(\vec{w}) \left( \prod_{k=1}^K p(v_k | \vec{v}_{\nb_{\G}(V_k)}^*, \vec{w}_{\pa_{\G}(V_k)}) \right) }{\prod_{k=1}^K p(v_k^* | \vec{v}_{\nb_{\G}(V_k)}^*, \vec{w}_{\pa_{\G}(V_k)}) } \right) \\
&\hspace{0.5cm}
\times \prod_{\vec{C} \in {\cal C}(\G_{\vec{V}}); |\vec{C}|\geq 2} \phi_{\vec{C}}(\vec{v}_{\vec{C}},\vec{w}_{\vec{C}^*}),
\label{eqn:lauritzen-cmrf-f}
\end{align}
where $\nb_{\G}(V_k)$ is the set of vertices in $\G$ with adjacencies to $V_k$ via undirected edges, $\vec{C}^* = \bigcap_{C \in \vec{C}} \pa_{\G}(C)$ for every $\vec{C}$, and $\phi_{\vec{C}}$ are terms defined as in (\ref{eqn:phi}).
Furthermore, all terms
$p(v_k | \vec{v}_{\nb_{\G}(V_k)}^*, \vec{w}_{\pa_{\G}(V_k)})$ (for all $k$) and $\phi_{\vec{C}}(\vec{v}_{\vec{C}},\vec{w}_{\vec{C}^*})$ for all $\vec{C} \in {\cal G}(\G_{\vec{V}})$ where $|\vec{C}| \geq 2$ in (\ref{eqn:lauritzen-cmrf-f}) are non-redundant and
variationally independent.
\label{cor:chen-lauritzen-cmrf}
\end{cor}
\begin{prf}
This follows immediately by Theorem \ref{thm:chen-lauritzen} and Theorem \ref{thm:h-c-cmrf}.
\end{prf}

In other words, (\ref{eqn:lauritzen-cmrf-f}) is obtained from (\ref{eqn:lauritzen-f}) due to the fact that any term in (\ref{eqn:lauritzen-f})
that does not correspond to a clique in $\G_{\vec{V}}$ becomes trivial (equal to $1$) whenever $p(\vec{v} | \vec{w})$ is in a CMRF model of a CUG $\G(\vec{V},\vec{W})$ (or an MRF model of a UG $\G(\vec{V})$ by taking $\vec{W}=\emptyset$).

Corollary \ref{cor:chen-lauritzen-cmrf} allows a non-redundant likelihood specification for parametric CMRF (or MRF) models that do not correspond to categorical or Gaussian data.  In addition, flexible semi-parametric likelihoods for CMRFs (or MRFs) can be specified as well.  However, while the terms $p(v_k | \vec{v}_{\nb_{\G}(V_k)}^*, \vec{w}_{\pa_{\G}(V_k)})$ and $\phi_{\vec{C}}(\vec{v}_{\vec{C}},\vec{w}_{\vec{V}^*})$ can be specified independently of each other, the fact that they are specific closed-form functionals of the distribution $p(\vec{v} | \vec{w})$ implies that they must satisfy certain restrictions in order for the overall model likelihood to be well-defined.  
The restrictions on terms in (\ref{eqn:lauritzen-cmrf-f}) are given by the following result.

\begin{lem}
For any positive $p(\vec{v}|\vec{w})$ in the CMRF model of a CUG $\G(\vec{V},\vec{W})$,
\begin{itemize}
\item for any $V_k \in \vec{V}$, the term $p(v_k | \vec{v}_{\nb_{\G}(V_k)}^*,\! \vec{w}_{\pa_{\G}(V_k)})$ in (\ref{eqn:lauritzen-f}) must be non-negative, and integrate to $1$, 
\item for any $\vec{C} \subseteq \vec{V}$ such that $\vec{C} \in {\cal C}(\G_{\vec{V}})$ and $|\vec{C}| \geq 2$,
the term $\phi_{\vec{C}}(\vec{v}_{\vec{C}},\vec{w}_{\vec{C}^*})$ in (\ref{eqn:lauritzen-f})
must be non-negative and satisfy the following:
\begin{align}
(\forall V_i \in \vec{C}, \vec{v}_{\vec{C}\setminus\{V_i\}},\vec{w}_{\vec{C}^*}) \hspace{0.2cm} \phi_{\vec{C}}(v_i^*,\vec{v}_{\vec{C}\setminus\{V_i\}},\vec{w}_{\vec{C}^*}) = 1.
\label{eqn:restriction}
\end{align}
\end{itemize}
The terms are otherwise unrestricted.
\label{lem:conditions}
\end{lem}
\begin{prf}
The condition on terms $p(v_k | \vec{v}_{\nb_{\G}(V_k)}^*, \vec{w}_{\pa_{\G}(V_k)})$ follows since they are conditional probabilities, and on
terms $\phi_{\{ V_i,V_j \}}(\vec{v}_{\{ V_i,V_j \}},\vec{w}_{\vec{C}^*})$ follows by properties of conditional odds ratio functions.

Assume, by induction, the result holds for all $\phi_{\vec{C}}(\vec{v}_{\vec{C}},\vec{w}_{\vec{C}^*})$, where $|\vec{C}|=i$.
To establish the conclusion we show that for all $\phi_{\vec{D}}(\vec{v}_{\vec{D}},\vec{w}_{\vec{D}^*})$, where
$\vec{D}=\vec{C} \cup \{ V_i \}$, and for all $V_i \in \vec{D}$,
$\phi_{\vec{D}}(\vec{v}_{\vec{D}}, \vec{w}_{\vec{D}^*})
=
\frac{
\phi_{\vec{C}}(\vec{v}_{\vec{C}}, v_i, \vec{w}_{\vec{C}^*})
}{
\phi_{\vec{C}}(\vec{v}_{\vec{C}}, v_i^*, \vec{w}_{\vec{C}^*})
}$.
We have:
\begin{align*}
&\frac{
\phi_{\vec{C}}(\vec{v}_{\vec{C}}, v_i, \vec{w}_{\vec{C}^*})
}{
\phi_{\vec{C}}(\vec{v}_{\vec{C}}, v_i^*, \vec{w}_{\vec{C}^*})
}=
\end{align*}
\begin{align*}
&=
\frac{
\exp \left\{
\sum_{\vec{B} \subseteq \vec{C}} (-1)^{|\vec{C} \setminus \vec{B}|} H_{\vec{B}}(\vec{c}_{\vec{B} \cup \{ V_i \}}, \vec{w}_{\vec{C}^*})
\right\}
}{
\exp \left\{
\sum_{\vec{B} \subseteq \vec{C}} (-1)^{|\vec{C} \setminus \vec{B}|} H_{\vec{B}}(\vec{c}_{\vec{B}}, \vec{w}_{\vec{C}^*})
\right\}
}\\
&=
\frac{
\exp \left\{
\sum_{\{ R_i \} \subseteq \vec{B} \subseteq \vec{D}} (-1)^{|\vec{D} \setminus \vec{B}|} H_{\vec{B}}(\vec{c}_{\vec{B}}, \vec{w}_{\vec{C}^*})
\right\}
}{
\exp \left\{
\sum_{\{R_i\} \not\subseteq \vec{B} \subseteq \vec{D}} (-1)^{|\vec{D} \setminus \vec{B}|-1} H_{\vec{B}}(\vec{c}_{\vec{B}}, \vec{w}_{\vec{C}^*})
\right\}
}\\
&=
\exp \left\{
\sum_{\vec{B} \subseteq \vec{D}} (-1)^{|\vec{D} \setminus \vec{B}|} H_{\vec{B}}(\vec{c}_{\vec{B}}, \vec{w}_{\vec{C}^*})
\right\}
= \phi_{\vec{D}}(\vec{v}_{\vec{D}}, \vec{w}_{\vec{C}^*}).
\end{align*}
That $\phi_{\vec{D}}(\vec{v}_{\vec{D}}, \vec{w}_{\vec{C}^*})$ is only a function of $\vec{v}_{\vec{D}}$ and $\vec{w}_{\vec{D}^*}$ (note that $\vec{D}^* \subseteq \vec{C}^*$ by definition) follows since
$p(\vec{v}|\vec{w})$ is in the CMRF model for $\G(\vec{V},\vec{W})$, and Theorem \ref{thm:h-c-cmrf}.
\end{prf}

{There are two ways to view this result.  If we are given a conditional distribution $p(\vec{v} | \vec{w})$ in the CMRF model for ${\cal G}(\vec{V},\vec{W})$, Lemma~\ref{lem:conditions} will tell us the properties that pieces of the L-C factorization of $p(\vec{v}|\vec{w})$ must satisfy.  Conversely, if we are interested in constructing a conditional distribution that lies in the CMRF model for a particular ${\cal G}(\vec{V},\vec{W})$, Lemma~\ref{lem:conditions} will tell us what restrictions to impose on pieces of an L-C factorization such that they form an element of the model.}
See also Lemmas 1 and 2 in prior work by \cite{chen10compatibility}.

\section{LIKELIHOODS FOR BAYESIAN NETWORK MODELS}
\label{sec:dags}

Bayesian network (BN) models admit natural likelihoods which parameterize each Markov factor $p(\vec{v}_{\{V\}} | \vec{v}_{\pa_{\G}(V)})$.  
Two distinct DAGs $\G_1$ and $\G_2$ may imply the same BN model in the sense that d-separation statements in $\G_1$ and $\G_2$ imply the same list of conditional independences.  In this case, DAGs $\G_1$ and $\G_2$ are said to lie in the same Markov equivalence class.  An elegant result states that $\G_1$ and $\G_2$ are equivalent in this sense if and only if they share edge adjacencies, and the same unshielded colliders (vertex triplets of the form $A \to C \gets B$, where $A$ and $B$ are not adjacent) \citep{verma90equiv}.

Likelihoods that parameterize $p(\vec{v}_{\{V\}} | \vec{v}_{\pa_{\G}(V)})$ terms for Gaussian or categorical data are coherent for the Markov structure
implied by the BN model in the sense that they will coincide for distinct DAGs within the Markov equivalence class.  However, general likelihoods do not have this property. which complicates their use in applications where this  coherence with respect to Markov structure is desirable, such as score based model selection algorithms that use data to select the best Markov model, or efficient semi-parametric estimators, with efficiency gains implied by Markov restrictions imposed on the data.  \cite{huang18generalized} provide an excellent discussion of this issue in the context of model selection.

We show how to use the Lauritzen-Chen likelihood to derive general parametric and semi-parametric likelihoods for DAG models that coincide within the Markov equivalence class.  We will do so by imposing the likelihood on a special type of mixed graph called the \emph{essential graph} \citep{andersson97characterization}, which represents the Markov equivalent class of DAGs.  Essential graphs are a special case of mixed graph graphical models known as chain graph models \citep{lauritzen96graphical}, and we will derive likelihoods for them first.  We will restriction attention to chain graph models under the Lauritzen-Wermuth-Frydenberg (LWF) interpretation \citep{lauritzen96graphical}.

A chain graph (CG) is a mixed graph containing directed and undirected edges with the property that no partially directed cycles are present.  A partially directed cycle is a sequence of consecutive edges {that contains at least one directed edge} with the property that undirected edges in this {sequence} can be oriented in such a way as to create a directed cycle.

Given a CG $\G$, a \emph{block} is an undirected connected component.  The set of blocks ${\cal B}(\G)$ in a CG $\G$ form a partition of vertices $\vec{V}$ in $\G$.
The CG model under the Lauritzen-Wermuth-Frydenberg (LWF) interpretation may be defined by the following factorization which generalizes the BN and MRF factorizations.
\begin{align}
\notag
p(\vec{v}) &= \prod_{\vec{B} \in {\cal B}(\G)} p(\vec{v}_{\vec{B}} | \vec{v}_{\pa_{\G}(\vec{B})})\\
&= \prod_{\vec{B} \in {\cal B}(\G)} \left( \prod_{\vec{C} \in {\cal C}(\G_{\vec{B}})} \phi_{\vec{C}}(\vec{v}_{\vec{C}},\vec{v}_{\vec{C}^*}) \right),
\label{eqn:cg-f}
\end{align}
where $\vec{C}^* = \bigcap_{C \in \vec{C}} \pa_{\G}(C)$ for every $\vec{C} \in \G_{\vec{B}}$ and $\vec{B} \in {\cal B}(\G)$.

In other words, the LWF CG factorization is a two level factorization.  The outer factorization of $p(\vec{v})$ resembles a DAG factorization, but formulated on elements of ${\cal B}(\G)$ rather than individual vertices. The inner factorization for every term $p(\vec{v}_{\vec{B}} | \vec{v}_{\pa_{\G}(\vec{B})})$ of the outer factorization is a CMRF factorization with respect to the conditional graph $\G(\vec{B},\pa_{\G}(\vec{B}))$ obtained from the CG $\G$ by restricting to vertices in $\vec{B} \cup \pa_{\G}(\vec{B})$, dropping all edges among $\pa_{\G}(\vec{B})$, and treating $\pa_{\G}(\vec{B})$ as fixed vertices.

This immediately implies the following result.
\begin{cor}
{\bf (Lauritzen-Chen (L-C) factorization of LWF chain graph models)}
For any positive $p(\vec{v})$ that factorizes with respect to a CG $\G$, $p(\vec{v})$ also obeys the following factorization:
\begin{align}
\notag
\prod_{\vec{B} \in {\cal B}(\G)}
\!\!\!
\frac{\phi_{\emptyset}(\vec{v}_{\pa_{\G}(\vec{B})})
\!\!
\prod\limits_{B \in \vec{B}} p(\vec{v}_{\{B\}} | \vec{v}_{\nb_{
	\G_{\vec{B}}
	}
	(B)}^*, \vec{v}_{\pa_{\G}(B)})
}{\prod\limits_{B \in \vec{B}} p(\vec{v}_{\{B\}}^* | \vec{v}_{\nb_{
	\G_{\vec{B}}
	}
	(B)}^*, \vec{v}_{\pa_{\G}(B)}) }
\\
\times \prod_{\vec{C} \in {\cal C}(\G_{\vec{B}}); |\vec{C}|\geq 2} \phi_{\vec{C}}(\vec{v}_{\vec{C}},\vec{v}_{\vec{C}^*})
\label{eqn:lauritzen-cg-f}
\end{align}
where for every $\vec{B} \in {\cal B}(\G)$, and every $\vec{C} \in {\cal C}(\G_{\vec{B}})$, the term $\phi_{\vec{C}}(\vec{v}_{\vec{C}},\vec{v}_{\vec{C}^*})$ is defined as in (\ref{eqn:moebius-terms}) and 
(\ref{eqn:phi})
using
$H_{\vec{D}}(\vec{v}_{\vec{D}}, \vec{v}_{\vec{C}^*}) = \log p(\vec{v}_{\vec{D}}, \vec{v}^*_{\vec{B} \setminus \vec{D}} | \vec{v}_{\pa_{\G}(\vec{B})})$ for every $\vec{D} \subseteq \vec{C}$.

Furthermore, all terms $\{ \phi_{\vec{C}}(\vec{v}_{\vec{C},\vec{v}_{\vec{C}^*}}) : \vec{C} \in {\cal C}(\G_{\vec{B}}), \vec{B} \in {\cal B}(\G) \}$, and
$\{ p(\vec{v}_{\{B\}} | \vec{v}_{\nb_{
	\G_{\vec{B}}
	}
	(B)}^*, \vec{v}_{\pa_{\G}(B)})  : B \in \vec{B} \in {\cal B}(\G) \}$
in (\ref{eqn:lauritzen-cg-f}) are non-redundant and variationally independent.
\label{cor:chen-lauritzen-cg-f}
\end{cor}
\begin{prf}
This follows from the fact that the inner LWF CG factorization is a CMRF factorization, Corollary \ref{cor:chen-lauritzen-cmrf}, and that terms in the outer CG factorization are non-redundant and variationally independent.
\end{prf}

Given a DAG $\G$, an \emph{essential graph} $\G^e$ is a mixed graph with a directed edge $V_i \to V_j$ whenever every DAG in the Markov equivalence class contains the edge $V_i \to V_j$, and an undirected edge $V_i - V_j$ whenever $V_i$ and $V_j$ are adjacent in every DAG in the Markov equivalence class, but the orientation of the edge differs for different elements of the equivalence class.  Results by \cite{andersson97characterization} show that the essential graph $\G^e$ is a CG, and give a polynomial time algorithm for its construction, given any DAG $\G$.

The following result is an immediate consequence of the results derived by \cite{andersson97characterization}.
\begin{prop}
For any DAG $\G$, a distribution $p(\vec{v})$ obeys the DAG factorization respect to $\G$ if and only if $p(\vec{v})$ obeys the LWF CG factorization with respect to $\G^e$.
\label{prop:essential}
\end{prop}

Proposition \ref{prop:essential} and Corollary \ref{cor:chen-lauritzen-cg-f} immediately yield coherent 
likelihoods for DAG models that are guaranteed to coincide for distinct DAGs in the same equivalence class.  {As was the case for MRFs, these likelihoods can be specified for both parametric and semi-parametric models.}
Such likelihoods have previously been derived in the special case of categorical data \citep{castelo04learning}.

\begin{figure*}
	\begin{center}
		\begin{tikzpicture}[>=stealth, node distance=1.0cm]
		\tikzstyle{format} = [draw, thick, circle, minimum size=4.0mm,
		inner sep=1pt]
		\tikzstyle{unode} = [draw, thick, circle, minimum size=1.0mm,
		inner sep=0pt,outer sep=0.9pt]
		\tikzstyle{square} = [draw, very thick, rectangle, minimum size=4mm]

	\begin{scope}[xshift=0cm]
		\path[-,  line width=0.9pt]
		node[format, shape=ellipse] (a) {$A$}
		node[format, shape=ellipse, right of=a, yshift=-0.0cm] (b) {$B$}			
		node[format, shape=ellipse, below of=a, yshift=+0.0cm] (c) {$C$}
		node[format, shape=ellipse, right of=c, xshift=0.0cm] (d) {$D$}

		(a) edge (b)
		(a) edge (c)
		(b) edge (d)
		(c) edge (d)

		node[below of=c, xshift=0.5cm, yshift=0.5cm]{(a)}	
		;
	\end{scope}

	\begin{scope}[xshift=2.3cm]
		\path[->,  line width=0.9pt]
		node[format, shape=ellipse, yshift=-0.0cm] (b) {$B$}			
		node[format, shape=ellipse, below of=b, yshift=+0.0cm] (c) {$C$}
		node[format, shape=ellipse, right of=b, xshift=0.0cm] (d) {$D$}
		node[format, shape=ellipse, right of=c, xshift=0.0cm] (e) {$E$}

		node[format, shape=ellipse, left of=b, xshift=0.4cm, yshift=-0.5cm] (a) {$A$}
		node[format, shape=ellipse, right of=d, xshift=-0.4cm, yshift=-0.5cm] (f) {$F$}

		(a) edge[blue] (b)
		(a) edge[blue] (c)
		(b) edge[blue] (d)
		(c) edge[blue] (d)
		(b) edge[blue] (e)
		(c) edge[blue] (e)
		(d) edge[blue] (e)

		(d) edge[blue] (f)
		(e) edge[blue] (f)
		(b) edge[blue] (f)
		(c) edge[blue] (f)

		node[below of=c, xshift=0.5cm, yshift=0.5cm]{(b)}	
		;
	\end{scope}

	\begin{scope}[xshift=5.3cm]
		\path[->,  line width=0.9pt]
		node[format, shape=ellipse, yshift=-0.0cm] (b) {$B$}			
		node[format, shape=ellipse, below of=b, yshift=+0.0cm] (c) {$C$}
		node[format, shape=ellipse, right of=b, xshift=0.0cm] (d) {$D$}
		node[format, shape=ellipse, right of=c, xshift=0.0cm] (e) {$E$}

		node[format, shape=ellipse, left of=b, xshift=0.4cm, yshift=-0.5cm] (a) {$A$}
		node[format, shape=ellipse, right of=d, xshift=-0.4cm, yshift=-0.5cm] (f) {$F$}

		(b) edge[blue] (a)
		(a) edge[blue] (c)
		(b) edge[blue] (d)
		(c) edge[blue] (d)
		(b) edge[blue] (e)
		(c) edge[blue] (e)
		(e) edge[blue] (d)

		(f) edge[blue] (d)
		(f) edge[blue] (e)
		(b) edge[blue] (f)
		(c) edge[blue] (f)

		node[below of=c, xshift=0.5cm, yshift=0.5cm]{(c)}	
		;
	\end{scope}

	\begin{scope}[xshift=8.3cm]
		\path[->,  line width=0.9pt]
		node[format, shape=ellipse, yshift=-0.0cm] (b) {$B$}			
		node[format, shape=ellipse, below of=b, yshift=+0.0cm] (c) {$C$}
		node[format, shape=ellipse, right of=b, xshift=0.0cm] (d) {$D$}
		node[format, shape=ellipse, right of=c, xshift=0.0cm] (e) {$E$}

		node[format, shape=ellipse, left of=b, xshift=0.4cm, yshift=-0.5cm] (a) {$A$}
		node[format, shape=ellipse, right of=d, xshift=-0.4cm, yshift=-0.5cm] (f) {$F$}

		(a) edge[blue] (b)
		(c) edge[blue] (a)
		(b) edge[blue] (d)
		(c) edge[blue] (d)
		(b) edge[blue] (e)
		(c) edge[blue] (e)
		(e) edge[blue] (d)

		(f) edge[blue] (d)
		(f) edge[blue] (e)
		(b) edge[blue] (f)
		(c) edge[blue] (f)

		node[below of=c, xshift=0.5cm, yshift=0.5cm]{(d)}	
		;
	\end{scope}
	
	\begin{scope}[xshift=11.3cm]
		\path[->,  line width=0.9pt]
		node[format, shape=ellipse, yshift=-0.0cm] (b) {$B$}			
		node[format, shape=ellipse, below of=b, yshift=+0.0cm] (c) {$C$}
		node[format, shape=ellipse, right of=b, xshift=0.0cm] (d) {$D$}
		node[format, shape=ellipse, right of=c, xshift=0.0cm] (e) {$E$}

		node[format, shape=ellipse, left of=b, xshift=0.4cm, yshift=-0.5cm] (a) {$A$}
		node[format, shape=ellipse, right of=d, xshift=-0.4cm, yshift=-0.5cm] (f) {$F$}

		(a) edge[-] (b)
		(c) edge[-] (a)
		(b) edge[blue] (d)
		(c) edge[blue] (d)
		(b) edge[blue] (e)
		(c) edge[blue] (e)
		(e) edge[-] (d)

		(f) edge[-] (d)
		(f) edge[-] (e)
		(b) edge[blue] (f)
		(c) edge[blue] (f)

		node[below of=c, xshift=0.5cm, yshift=0.5cm]{(e)}	
		;
	\end{scope}
		
		\end{tikzpicture}
	\end{center}
	\caption{
	(a) An MRF model with 
	two restrictions: $(A \ci D | B,C)$, and $(B \ci C | A,D)$.
	(b), (c), (d) Three distinct DAGs in the same Markov equivalence class corresponding to the Bayesian network (BN) model with restrictions $(B \ci C | A)$, $(D,E,F \ci A | B,C)$.
	(e) The essential graph CG corresponding to the Markov equivalence class containing DAGs in (b), (c), and (d).
	}
	\label{fig:examples}
\end{figure*}
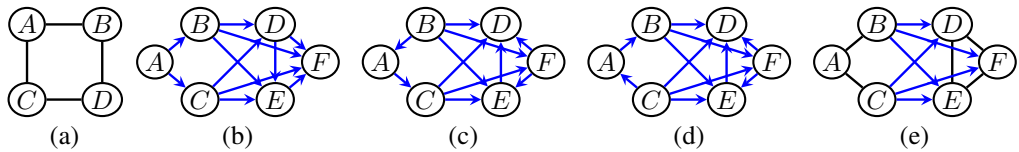

\section{EXAMPLES}


Figure~\ref{fig:examples} (a) shows an MRF with 
two restrictions: $(A \ci D | B,C)$, and $(B \ci C | A,D)$.  The L-C factorization of any positive distribution $p(a,b,c,d)$ in this MRF model is:
\begin{align*}
& Z^{-1} p(a | b^*,c^*) p(b | a^*,d^*) p(c | a^*,d^*) p(d | b^*,c^*) \\
&\hspace{0.5cm}
\underbrace{OR(a, b | c^*,d^*)}_{\phi_{\{A,B\}}} \underbrace{OR(a, c | b^*,d^*)}_{\phi_{\{A,C\}}} \underbrace{OR(b, d | a^*,c^*)}_{\phi_{\{B,D\}}} 
\\
&\hspace{0.5cm}
\underbrace{OR(c,d | a^*,b^*)}_{\phi_{\{C,D\}}}
\end{align*}
where
\begin{align*}
Z = \frac{p(a^* | b^*,c^*) p(b^* | a^*,d^*)
p(c^* | a^*,d^*) p(d^* | b^*,c^*)}{p(a^*, b^*, c^*, d^*)}.
\end{align*}
Note that
there 
are no terms corresponding to any subset of $\{ A, B, C, D \}$ of size $3$ or $4$, as no such set corresponds to 
a clique in the graph in Figure~\ref{fig:examples} (a).  
This is the simplest MRF model not observationally equivalent to any DAG model.  This model may be generalized to MRFs with graphs forming a square grid.  Square grid MRFs have been used as spin models in physics \citep{nishimori01statistical}, the original application of MRF models due to \cite{ising25beitrag}, and in computer vision to model images composed of pixels \citep{szeliski22computer}.
An example of such an MRF 
with the corresponding factorization is given in the Appendix.

Figures~\ref{fig:examples} (b), (c), and (d) show three (out of eighteen) distinct DAGs that are in the same Markov equivalence class, and thus imply the same BN model.  The essential graph corresponding to this Markov equivalence class is shown in Figure~\ref{fig:examples} (e).  Note that all edges that differ in their orientation in Figures~\ref{fig:examples} (b), (c), and (d) are represented by undirected edges in Figure~\ref{fig:examples} (e).
The factorization for $p(a,b,c,d,e,f)$ associated with Figure~\ref{fig:examples} (e) yields a likelihood that obeys Markov restrictions in the equivalence class, and coincides for all DAGs in the class by construction.  It is given as follows:
\begin{align*}
&
Z_1^{-1}
p(a | b^*, c^*) p(c | a^*) p(b | a^*) OR(a,b | c^*) OR(a,c | b^*)
\times\\
&
Z_2(b,c)^{-1}
p(d | b,c,e^*,\!f^*) p(e | b,c,d^*\!,\!f^*) p(f | b,c,d^*\!,e^*)\times\\
& OR(d,e | b,c,f^*) OR(d,f | b,c,e^*) OR(e,f | b,c,d^*) \times\\
& \frac{ OR(d,e | f, b,c) }{ OR(d,e | \!f^*\!, b,c) }.
\end{align*}

Additional examples of the L-C factorization in more complex models, arising in areas such as philogenetic analysis and computer vision, are discussed in the Appendix.



\section{DISCUSSION}
\label{sec:discussion}


In this paper we have shown that decompositions of joint distributions described by 
\cite{chen03note,chen07semiparametric,chen10compatibility} and 
\cite{lauritzen96graphical} are closely connected.  This allowed us to show that the clique factorization for Markov random field graphical models may be specified using univariate conditional distribution terms, and alternate product terms that arise from the M\"obius inversion formula and that generalize the odds ratio function.  We show that these terms are \emph{variationally independent}, and \emph{non-redundant}. 
This result is also generalized to conditional Markov random field models, directed acyclic graph (DAG) models, and chain graph (CG) models.  In particular, these results allow parametric and semi-parametric likelihoods for DAG models to be specified that coincide, by construction, among all elements of the Markov equivalence class of any DAG model.  This solves a long-standing open issue in the score based causal discovery literature
\citep{huang18generalized}.

The proposed MRF likelihoods may be viewed as a 
generalization of hierarchical log-linear models for categorical data \citep{lauritzen96graphical}, where univariate conditional terms $p(v_k | \vec{v}_{-k},\vec{w})$ generalize main effect parameters, and generalized odds ratio terms $\phi_{\vec{C}}(\vec{v}_{\vec{C}},\vec{w}_{\vec{C}^*})$ generalize higher order ($|\vec{C}|$-way) interaction parameters in such models.

A natural semi-parametric modeling approach for likelihoods we describe is to use flexible methods for univariate conditional terms, such as those based on kernel regressions \citep{nadaraya64on,watson64smooth}, and parametric forms for generalized interaction terms, for example the transformed linear generalized odds ratio model: $\log \phi_{\vec{C}}(\vec{v}_{\vec{C}},\vec{w}_{\vec{C}^*}; \gamma) = \gamma \prod_{C \in \vec{C}} (g_C(\vec{v}_C, \vec{w}_{\vec{C}^*}) - g_C(\vec{v}_C^*, \vec{w}_{\vec{C}^*}))$ for fixed or unknown functions $g_C$.  Such likelihoods may be optimized by iterative methods similarly to other semi-parametric likelihood models, such as the projection pursuit model \citep{friedman81projection}.

We note that the proposed likelihood differs from copula models \citep{sklar59fonctions}.  The latter aims to link together univariate \emph{marginal} distribution by a (typically parametric) object called the \emph{copula} that captures joint behavior of the random variables in the model.  On the other hand, the univariate terms in the proposed likelihoods are univariate \emph{conditional} distributions, with the other terms capturing interactions among variables.

If the terms in the proposed likelihoods are kept unrestricted, aside from necessary restrictions imposed by Lemma \ref{lem:conditions}, the result yields a useful view on the tangent space of the corresponding Markov model, which is useful for deriving estimators based on influence functions that attain the semi-parametric efficiency bound.  Indeed, a special case of the Chen decomposition for a particular class of Markov random fields has already been used to derive an efficient influence function in a missing data model \citep{malinsky2021semiparametric}. 

{Additional work is needed to obtain practical estimators for graphical model parameters using} the proposed likelihoods.
{This is because the likelihoods we describe}
inherit the usual difficulties of known parametric likelihoods for MRFs and CMRFs: the computation of the normalizing constant $Z$ or normalizing function $Z(\vec{w})$.  Existing approaches to this problem include approximation of these objects by numerical integration, sum-product algorithms in sufficiently sparse graphs \citep{koller09probabilistic}, variational inference \citep{wainwright08graphical}, or reformulation of the problem via composite likelihoods \citep{varin11overview} where the need for normalization disappears.  For a review of these methods, see \citep{wainwright08graphical}.
Application of these approaches to the proposed likelihoods is an interesting area of future work.  In particular, the application of sum-product algorithms would entail developing classes of likelihoods where intermediate objects obtained by marginalization and products of clique factors maintain a particular form.

\subsubsection*{Acknowledgements}
The author would like to thank reviewers for their helpful comments which led to a much improved manuscript.
The author would like to gratefully acknowledge the Simons Institute program on Causality in the Spring of 2022 for providing inspiration for this work.
The author was supported in part by the ONR grant N00014-21-1-2820,
NSF grant 2040804, NSF CAREER grant 1942239, and the NIH grant R01 AI127271-01A1.

\bibliography{references}

\appendix
\onecolumn

\section{PROOFS}
\label{supp:proofs}

\begin{proa}{\ref{prop:chen}}
The joint distribution $p(v_1, v_2 | \vec{w})$ admits the following decomposition.
\begin{align*}
p(v_1, v_2 | \vec{w}) =
\frac{
p(v_1 | v_2^*, \vec{w}) p(v_2 | v_1^*, \vec{w}) OR(v_1, v_2 | \vec{w})
}{
Z(\vec{w})
}
\end{align*}
Moreover, all terms in the numerator of this decomposition are \emph{non-redundant} and \emph{variationally independent}.
\end{proa}

\begin{prf}
Here we follow the structure of the arguments presented by \cite{chen03note}, as well as the main body and the Appendix of
the paper by \cite{chen07semiparametric}.  We generalize the argument of \cite{chen03note} somewhat to apply to conditional distributions.
In the following, summation may be replaced by integration where appropriate.

We will first show that
\begin{align*}
p(v_1 | v_2,\vec{w}) &= \frac{ p(v_1 | v_2^*,\vec{w}) OR(v_1, v_2 | \vec{w}) }{ \sum_{v_1} p(v_1 | v_2^*,\vec{w}) OR(v_1,v_2|\vec{w}) },\\
p(v_2 | v_1,\vec{w}) &= \frac{ p(v_2 | v_1^*,\vec{w}) OR(v_1, v_2 | \vec{w}) }{ \sum_{v_2} p(v_2 | v_1^*,\vec{w}) OR(v_1,v_2|\vec{w}) }.
\end{align*}
Next, we will show that $p(v_1 | v_2^*,\vec{w})$ may be specified in a non-redundant and variationally independent way from any function of $v_2$ and $\vec{w}$ and in particular from $p(v_2 | v_1^*,\vec{w})$.
Symmetrically we will show that $p(v_2 | v_1^*,\vec{w})$ may be specified in a non-redundant and variationally independent way from any function of $v_1$ and $\vec{w}$ and in particular from $p(v_1 | v_2^*,\vec{w})$.
Finally, we will derive that the numerator of the decomposition are non-redundant and variationally independent.

We start by noting that by Bayes rule, $p(v_1 | v_2^*, \vec{w}) = \frac{ p(v_2^* | v_1, \vec{w}) p(v_1 | \vec{w}) }{ \sum_{v_1} p(v_2^* | v_1, \vec{w}) p(v_1 | \vec{w}) }$.
This implies
\begin{align*}
p(v_1 | \vec{w})
&= \frac{ p(v_1 | v_2^*, \vec{w}) \sum_{v_1} p(v_2^* | v_1 | \vec{w}) p(v_1 | \vec{w}) }{ p(v_2^* | v_1, \vec{w}) }\\
&= \frac{ p(v_1 | v_2^*, \vec{w}) \sum_{v_1} p(v_2^* | v_1,\vec{w}) \left( p(v_1 | v_2^*, \vec{w}) p(v_2^* | \vec{w}) / p(v_2^* | v_1, \vec{w}) \right)  }{ p(v_2^* | v_1, \vec{w}) }\\
&= \frac{ p(v_1 | v_2^*,\vec{w}) }{ p(v_2^* | v_1,\vec{w}) } \sum_{v_1} p(v_1 | v_2^*, \vec{w}) p(v_2^*| \vec{w})\\
&= \frac{ p(v_1 | v_2^*,\vec{w}) }{ p(v_2^* | v_1,\vec{w}) } / \sum_{v_1} \frac{ p(v_1 | v_2^*,\vec{w}) }{ p(v_2^* | v_1,\vec{w}) },
\end{align*}
where the last equality follows from the fact that
\begin{align*}
\sum_{v_1} p(v_1 | v_2^*,\vec{w}) p(v_2^* | \vec{w}) \times \sum_{v_1} \frac{p(v_1 | v_2^*,\vec{w})}{p(v_2^* | v_1)} &= \sum_{v_1} \frac{p(v_1 | v_2^*,\vec{w}) p(v_2^* | \vec{w})}{ p(v_2^* | v_1,\vec{w})} = \sum_{v_1} \frac{p(v_2^*,v_1 | \vec{w})}{ p(v_2^* | v_1,\vec{w})}
= \sum_{v_1} p(v_1 | \vec{w}) = 1.
\end{align*}

Applying Bayes rule again, we see that
\begin{align*}
p(v_1 | v_2, \vec{w}) &= \frac{ p(v_2 | v_1, \vec{w}) p(v_1 | \vec{w}) }{ \sum_{v_1} p(v_2 | v_1,\vec{w}) p(v_1 | \vec{w}) }
= \frac{ p(v_2 | v_1,\vec{w}) \frac{ p(v_1 | v_2^*,\vec{w}) }{ p(v_2^* | v_1,\vec{w}) } / \sum_{v_1} \frac{ p(v_1 | v_2^*,\vec{w}) }{ p(v_2^* | v_1,\vec{w}) } }{ \sum_{v_1} p(v_2 | v_1,\vec{w}) \frac{ p(v_1 | v_2^*,\vec{w}) }{ p(v_2^* | v_1,\vec{w}) } / \sum_{v_1} \frac{ p(v_1 | v_2^*,\vec{w}) }{ p(v_2^* | v_1,\vec{w}) } }
= \frac{ p(v_2 | v_1,\vec{w}) \frac{ p(v_1 | v_2^*,\vec{w}) }{ p(v_2^* | v_1,\vec{w}) } }{ \sum_{v_1} p(v_2 | v_1,\vec{w}) \frac{ p(v_1 | v_2^*,\vec{w}) }{ p(v_2^* | v_1,\vec{w}) } }.
\end{align*}

Since the conditional odds ratio function $OR(v_1, v_2; v_1^*, v_2^* | \vec{w})$ given reference values $v_1^*,v_2^*$ is defined as $\frac{ p(v_2 | v_1,\vec{w}) p(v_2^* | v_1^*,\vec{w})}{ p(v_2^* | v_1,\vec{w}) p(v_2 | v_1^*,\vec{w}) }$ (or equivalently as $\frac{ p(v_1 | v_2,\vec{w}) p(v_1^* | v_2^*,\vec{w})}{ p(v_1^* | v_2,\vec{w}) p(v_1 | v_2^*,\vec{w}) }$), we have:
\begin{align}
p(v_1 | v_2,\vec{w}) &= \frac{ p(v_2 | v_1,\vec{w}) \frac{ p(v_1 | v_2^*,\vec{w}) }{ p(v_2^* | v_1,\vec{w}) } }{ \sum_{v_1} p(v_2 | v_1,\vec{w}) \frac{ p(v_1 | v_2^*,\vec{w}) }{ p(v_2^* | v_1,\vec{w}) } }
= \frac{ p(v_2 | v_1,\vec{w}) \frac{ p(v_1 | v_2^*,\vec{w}) }{ p(v_2^* | v_1,\vec{w}) } \frac{ p(v_2^* | v_1^*,\vec{w}) }{ p(v_2 | v_1^*,\vec{w}) } }{ \frac{ p(v_2^* | v_1^*,\vec{w}) }{ p(v_2 | v_1^*,\vec{w}) } \sum_{v_1} p(v_2 | v_1,\vec{w}) \frac{ p(v_1 | v_2^*,\vec{w}) }{ p(v_2^* | v_1,\vec{w}) } }
= \frac{ p(v_1 | v_2^*,\vec{w}) OR(v_1, v_2 | \vec{w}) }{ \sum_{v_1} p(v_1 | v_2^*,\vec{w}) OR(v_1,v_2|\vec{w}) },
\label{eqn:1-given-2}
\end{align}
where we shorten the conditional odds ratio function notation to $OR(v_1, v_2 | \vec{w})$ by keeping the reference values implicit.
Since this argument is completely symmetric with respect to $v_1$ and $v_2$, we also have:
\begin{align}
p(v_2 | v_1,\vec{w}) &= \frac{ p(v_2 | v_1^*,\vec{w}) OR(v_1, v_2 | \vec{w}) }{ \sum_{v_2} p(v_2 | v_1^*,\vec{w}) OR(v_1,v_2|\vec{w}) }.
\label{eqn:2-given-1}
\end{align}

Next, note that (\ref{eqn:1-given-2}), (\ref{eqn:2-given-1}) and the chain rule of probability imply:
\begin{align}
p(v_1, v_2 | \vec{w})
&= \frac{ p(v_1 | v_2^*,\vec{w}) OR(v_1, v_2 | \vec{w}) }{ \sum_{v_1} p(v_1 | v_2^*,\vec{w}) OR(v_1,v_2|\vec{w}) } p(v_2 | \vec{w})
= \frac{ p(v_2 | v_1^*,\vec{w}) OR(v_1, v_2 | \vec{w}) }{ \sum_{v_2} p(v_2 | v_1^*,\vec{w}) OR(v_1,v_2|\vec{w}) } p(v_1 | \vec{w}).
\label{eqn:v1v2-rep}
\end{align}
Thus,
\begin{align}
p(v_1 | \vec{w}) = \sum_{v_2} p(v_1,v_2 | \vec{w}) = p(v_1 | v_2^*,\vec{w}) \sum_{v_2} \frac{ OR(v_1, v_2 | \vec{w}) }{ \sum_{v_1} p(v_1 | v_2^*,\vec{w}) OR(v_1,v_2|\vec{w}) } p(v_2 | \vec{w}).
\label{eqn:v1-rep}
\end{align}
Next, note that (\ref{eqn:v1v2-rep}) and (\ref{eqn:v1-rep}) imply:
\begin{align*}
\frac{ p(v_2 | v_1^*,\vec{w}) }{ \sum_{v_2} p(v_2 | v_1^*,\vec{w}) OR(v_1,v_2|\vec{w}) }
= \frac{ p(v_1 | v_2^*,\vec{w}) }{ \sum_{v_1} p(v_1 | v_2^*,\vec{w}) OR(v_1,v_2|\vec{w}) } \frac{ p(v_2 | \vec{w}) }{ p(v_1 | \vec{w}) }\\
= \frac{ p(v_2 | \vec{w}) } { \sum_{v_1} p(v_1 | v_2^*,\vec{w}) OR(v_1,v_2|\vec{w}) } / \sum_{v_2} \frac{ OR(v_1, v_2 | \vec{w}) p(v_2 | \vec{w}) }{ \sum_{v_1} p(v_1 | v_2^*,\vec{w}) OR(v_1,v_2|\vec{w}) }.
\end{align*}
This, in turn, implies that:
\begin{align}
p(v_2 | \vec{w}) &= p(v_2 | v_1^*,\vec{w}) \frac{ \sum_{v_1} p(v_1 | v_2^*,\vec{w}) OR(v_1,v_2|\vec{w}) }{ \sum_{v_2} p(v_2 | v_1^*,\vec{w}) OR(v_1,v_2|\vec{w}) } \sum_{v_2} \frac{ OR(v_1, v_2 | \vec{w}) p(v_2 | \vec{w}) }{ \sum_{v_1} p(v_1 | v_2^*,\vec{w}) OR(v_1,v_2|\vec{w}) }.
\label{eqn:v2-rep}
\end{align}
Since $\sum_{v_2} p(v_2 | \vec{w}) = 1$,
\begin{align}
\notag
& \sum_{v_2} p(v_2 | v_1^*,\vec{w}) \frac{ \sum_{v_1} p(v_1 | v_2^*,\vec{w}) OR(v_1,v_2|\vec{w}) }{ \sum_{v_2} p(v_2 | v_1^*,\vec{w}) OR(v_1,v_2|\vec{w}) } \sum_{v_2} \frac{ OR(v_1, v_2 | \vec{w}) p(v_2 | \vec{w}) }{ \sum_{v_1} p(v_1 | v_2^*,\vec{w}) OR(v_1,v_2|\vec{w}) } = 1\\
&\Rightarrow
\sum_{v_1,v_2} p(v_2 | v_1^*,\vec{w}) p(v_1 | v_2^*,\vec{w}) OR(v_1,v_2|\vec{w}) =
\frac{
\sum_{v_2} p(v_2 | v_1^*,\vec{w}) OR(v_1,v_2|\vec{w})
}{
\sum_{v_2} \frac{ OR(v_1, v_2 | \vec{w}) p(v_2 | \vec{w}) }{ \sum_{v_1} p(v_1 | v_2^*,\vec{w}) OR(v_1,v_2|\vec{w}) }
}.
\label{eqn:v2-norm}
\end{align}
Equations (\ref{eqn:v2-rep}) and (\ref{eqn:v2-norm}) together imply:
\begin{align}
p(v_2 | \vec{w}) &=
\frac{
p(v_2 | v_1^*,\vec{w}) \sum_{v_1} p(v_1 | v_2^*,\vec{w}) OR(v_1,v_2|\vec{w})
}{
\sum_{v_1,v_2} p(v_2 | v_1^*,\vec{w}) p(v_1 | v_2^*,\vec{w}) OR(v_1,v_2|\vec{w})
},
\label{eqn:v2-ratio}
\end{align}
Rearranging (\ref{eqn:v2-ratio}) yields:
\begin{align*}
p(v_2 | v_1^*, \vec{w}) &=
\frac{
p(v_2 | \vec{w}) \sum_{v_1,v_2} p(v_2 | v_1^*,\vec{w}) p(v_1 | v_2^*,\vec{w}) OR(v_1,v_2|\vec{w})
}{
\sum_{v_1} p(v_1 | v_2^*,\vec{w}) OR(v_1,v_2|\vec{w})
}.
\end{align*}
Since $\sum_{v_2} p(v_2 | v_1^*, \vec{w}) = 1$,
\begin{align}
\notag
& \sum_{v_2} \frac{
p(v_2 | \vec{w}) \sum_{v_1,v_2} p(v_2 | v_1^*,\vec{w}) p(v_1 | v_2^*,\vec{w}) OR(v_1,v_2|\vec{w})
}{
\sum_{v_1} p(v_1 | v_2^*,\vec{w}) OR(v_1,v_2|\vec{w})
} = 1\\
& \Rightarrow
\notag
\sum_{v_2}
\frac{
p(v_2 | \vec{w})
}{
\sum_{v_1} p(v_1 | v_2^*,\vec{w}) OR(v_1,v_2|\vec{w})
}
=
\frac{ 1 }{
\sum_{v_1,v_2} p(v_2 | v_1^*,\vec{w}) p(v_1 | v_2^*,\vec{w}) OR(v_1,v_2|\vec{w})
}\\
& \Rightarrow
p(v_2 | v_1^*, \vec{w}) =
\frac{
p(v_2 | \vec{w})
}{
\sum_{v_1} p(v_1 | v_2^*,\vec{w}) OR(v_1,v_2|\vec{w})
}
/
\sum_{v_2}
\frac{
p(v_2 | \vec{w})
}{
\sum_{v_1} p(v_1 | v_2^*,\vec{w}) OR(v_1,v_2|\vec{w})
}.
\label{eqn:v2-base-rep}
\end{align}

Equations (\ref{eqn:v2-ratio}) and (\ref{eqn:v2-base-rep}) imply that $p(v_2 | \vec{w})$ and $p(v_2 | v_1^*,\vec{w})$ are connected by an (algebraic) bijective mapping.
Thus, since models for $p(v_1 | v_2^*, \vec{w})$ and $p(v_2 | \vec{w})$ admit a non-redundant and variationally independent parameterization, then so do
$p(v_1 | v_2^*, \vec{w})$ and $p(v_2 | v_1^*, \vec{w})$.

A symmetric argument may be used to establish the following identities:
\begin{align}
p(v_1 | \vec{w}) &=
\frac{
p(v_1 | v_2^*,\vec{w}) \sum_{v_2} p(v_2 | v_1^*,\vec{w}) OR(v_1,v_2|\vec{w})
}{
\sum_{v_1,v_2} p(v_1 | v_2^*,\vec{w}) p(v_2 | v_1^*,\vec{w}) OR(v_1,v_2|\vec{w})
},
\label{eqn:v1-ratio}
\end{align}
and
\begin{align}
p(v_1 | v_2^*, \vec{w}) =
\frac{
p(v_1 | \vec{w})
}{
\sum_{v_2} p(v_2 | v_1^*,\vec{w}) OR(v_1,v_2|\vec{w})
}
/
\sum_{v_1}
\frac{
p(v_1 | \vec{w})
}{
\sum_{v_2} p(v_2 | v_1^*,\vec{w}) OR(v_1,v_2|\vec{w})
}.
\label{eqn:v1-base-rep}
\end{align}

Combining either (\ref{eqn:v1-ratio}) or (\ref{eqn:v2-ratio}) and (\ref{eqn:v1v2-rep}) yields:
\begin{align*}
p(v_1, v_2 | \vec{w})
&=
\frac{
p(v_1 | v_2^*, \vec{w}) p(v_2 | v_1^*, \vec{w}) OR(v_1, v_2 | \vec{w})
}{
\sum_{v_1,v_2}
p(v_1 | v_2^*, \vec{w}) p(v_2 | v_1^*, \vec{w}) OR(v_1, v_2 | \vec{w})
}\\
&=
\frac{
p(v_1 | v_2^*, \vec{w}) p(v_2 | v_1^*, \vec{w}) OR(v_1, v_2 | \vec{w})
}{
Z(\vec{w})
}.
\end{align*}

Finally, we note that $p(v_1 | v_2, \vec{w})$ and $p(v_2 | \vec{w})$ admit a non-redundant and variationally independent parameterization.
However, (\ref{eqn:1-given-2}) suggests $p(v_1 | v_2, \vec{w})$ may be parameterized by $p(v_1 | v_2^*, \vec{w})$ and $OR(v_1, v_2 | \vec{w})$, and above argument suggests $p(v_2 | \vec{w})$ has a bijective map with $p(v_2 | v_1^*, \vec{w})$.  Thus, $p(v_2 | v_1^*, \vec{w})$ admits a parameterization that is non-redundant and variationally independent for $p(v_1 | v_2^*, \vec{w})$ and $OR(v_1, v_2 | \vec{w})$.
A symmetric argument implies $p(v_1 | v_2^*, \vec{w})$  admits a parameterization that is non-redundant and variationally independent for $p(v_2 | v_1^*, \vec{w})$ and $OR(v_1, v_2 | \vec{w})$.
Thus, all terms in the numerator of the decomposition of $p(v_1, v_2 | \vec{w})$ are non-redundant and variationally independent.
\end{prf}


\begin{lem}
$OR(v_3, (v_1,v_3)) = OR(v_1, v_3 | v_2^*)
OR(v_2, v_3 | v_1^*)
\frac{
OR(v_1, v_2 | v_3)
}{
OR(v_1, v_2 | v_3^*)
}$
\end{lem}
\begin{prf}
The conclusion follows by definition and term cancellation, as follows:
\begin{align*}
OR(v_3, (v_1, v_2))
&=
\frac{
p(v_3, (v_1, v_2)) p(v_3^*, (v_1^*, v_2^*))
}{
p(v_3^*, (v_1, v_2)) p(v_3, (v_1^*, v_2^*))
}
\\
OR(v_1, v_3 | v_2^*)
&=
\frac{
p(v_1, v_3, v_2^*) p(v_1^*, v_3^*, v_2^*) 
}{
p(v_1^*, v_3, v_2^*) p(v_1, v_3^*, v_2^*)
}
\\
OR(v_2, v_3 | v_1^*)
&=
\frac{
p(v_2, v_3, v_1^*) p(v_2^*, v_3^*, v_1^*) 
}{
p(v_2^*, v_3, v_1^*) p(v_2, v_3^*, v_1^*)
}
\\
\frac{
OR(v_1, v_2 | v_3)
}{
OR(v_1, v_2 | v_3^*)
}
&=
\frac{
	\frac{
		p(v_1, v_2, v_3) p(v_1^*, v_2^*, v_3)
	}{
		p(v_1^*, v_2, v_3) p(v_1, v_2^*, v_3)
	}
}{
	\frac{
		p(v_1, v_2, v_3^*) p(v_1^*, v_2^*, v_3^*)
	}{
		p(v_1^*, v_2, v_3^*) p(v_1, v_2^*, v_3^*)
	}
}
\\
OR(v_1, v_3 | v_2^*)
OR(v_2, v_3 | v_1^*)
\frac{
OR(v_1, v_2 | v_3)
}{
OR(v_1, v_2 | v_3^*)
}
&=
\frac{
\stkout{p(v_1, v_3, v_2^*)} \stkout{p(v_1^*, v_3^*, v_2^*)}
}{
\stkout{p(v_1^*, v_3, v_2^*)} \stkout{p(v_1, v_3^*, v_2^*)}
}
\frac{
\stkout{p(v_2, v_3, v_1^*)} \stkout{p(v_2^*, v_3^*, v_1^*) }
}{
\stkout{p(v_2^*, v_3, v_1^*)} \stkout{p(v_2, v_3^*, v_1^*)}
} 
\\
& \hspace{0.5cm}
\times
\frac{
	\frac{
		{\color{red}p(v_1, v_2, v_3)} {\color{red}p(v_1^*, v_2^*, v_3)}
	}{
		\stkout{p(v_1^*, v_2, v_3)} \stkout{p(v_1, v_2^*, v_3)}
	}
}{
	\frac{
		{\color{red}p(v_1, v_2, v_3^*)} {\color{red}p(v_1^*, v_2^*, v_3^*)}
	}{
		\stkout{p(v_1^*, v_2, v_3^*)} \stkout{p(v_1, v_2^*, v_3^*)}
	}
}\\
&=
{\color{red}
\frac{
p(v_3, (v_1, v_2)) p(v_3^*, (v_1^*, v_2^*))
}{
p(v_3^*, (v_1, v_2)) p(v_3, (v_1^*, v_2^*))
}
}
=
OR(v_3, (v_1, v_2))
\end{align*}
\end{prf}

\begin{thma}{\ref{thm:h-c-cmrf}}{\bf (Hammersly-Clifford for conditional MRFs)}
Assume a positive $p(\vec{v} | \vec{w})$ obeys the pairwise Markov property for a CUG $\G(\vec{V},\vec{W})$.
That is, for every $V \in \vec{V}$, $Z \in \vec{V} \cup \vec{W}$ such that $Z$ is non-adjacent to $V$ in $\G(\vec{V},\vec{W})$,
$p(v | (\vec{v} \cup \vec{w}) \setminus \{ v \})$ is only a function of $(\vec{v} \cup \vec{w}) \setminus \{ z \}$.
Then $p(\vec{v} | \vec{w})$ factorizes with respect to $\G(\vec{V},\vec{W})$.
That is, 
\begin{align*}
p(\vec{v} | \vec{w})
=
\prod_{\vec{C} \subseteq \bar{\cal C}(\G_{\vec{V}})} \phi_{\vec{C}}(\vec{v}_{\vec{C}}, \vec{w}_{\vec{C}^*}),
\end{align*}
where for every $\vec{C}$, $\vec{C}^* = \bigcap_{C \in \vec{C}} \pa_{\G}(C)$, and $\phi_{\vec{C}}$ are terms defined as in (\ref{eqn:moebius-inversion}).
\end{thma}

\begin{prf}
We generalize the structure of the proof for the MRF case found in the book by \cite{lauritzen96graphical} to CMRFs.

Fix distinct $Y,Z \in \vec{A} \subseteq \vec{V}$ that are not neighbors in ${\cal G}(\vec{V},\vec{W})$, and let $\vec{C} = \vec{A} \setminus \{ Y, Z \}$, 
Since  $p(\vec{v} | \vec{w})$ is pairwise Markov with respect to ${\cal G}(\vec{V},\vec{W})$, we have that $\tilde{\phi}_{\vec{A}}(\vec{v}_{\vec{A}},\vec{w})$ is equal to
\begin{align}
\notag
\sum_{\vec{b} : \vec{b} \subseteq \vec{c}} (-1)^{|\vec{c} \setminus \vec{b}|}
&
\left\{ H_{\vec{B}}(\vec{v}_{\vec{B}},\vec{w}) - H_{\vec{B} \cup \{ Y \}}(\vec{v}_{\vec{B} \cup \{ Y \}},\vec{w})\right.\\ 
\phantom{
\left\{ H_{\vec{B}}(\vec{v}_{\vec{B}},\vec{w})
\right\}.}
& \left. - H_{\vec{B} \cup \{ Z \}}(\vec{v}_{\vec{B} \cup \{ Z \}},\vec{w}) + H_{\vec{B} \cup \{ Y, Z \}}(\vec{v}_{\vec{B} \cup \{ Y, Z \}},\vec{w}) \right\}.
\label{eqn:pairwise-cancel}
\end{align}

Let $\vec{D} = \vec{V} \setminus \{ Y, Z \}$.  Then 
we have
{\small
\begin{align*}
& H_{\vec{B} \cup \{ Y, Z \}}(\vec{v}_{\vec{B} \cup \{ Y, Z \}},\vec{w}) - H_{\vec{B} \cup \{ Y \}}(\vec{v},\vec{w})\\
&=
\log \frac{
p(\vec{b},y,z,\vec{v}^*_{\vec{D} \setminus \vec{B}}|\vec{w})
}{
p(\vec{b},y,z^*,\vec{v}^*_{\vec{D} \setminus \vec{B}}|\vec{w})
} \text{ (by definition)}\\
&=
\frac{
p(y | \vec{b}, \vec{v}^*_{\vec{D} \setminus \vec{B}},\vec{w}) p(z | \vec{b}, \vec{v}^*_{\vec{D} \setminus \vec{B}},\vec{w})
}{
p(y | \vec{b}, \vec{v}^*_{\vec{D} \setminus \vec{B}},\vec{w}) p(z^* | \vec{b}, \vec{v}^*_{\vec{D} \setminus \vec{B}},\vec{w})
} \text{ ($Y \!\ci\! Z | \vec{D},\vec{W}$ by the pairwise property)}\\
&=
\frac{
p(y^* | \vec{b}, \vec{v}^*_{\vec{D} \setminus \vec{B}},\vec{w}) p(z | \vec{b}, \vec{v}^*_{\vec{D} \setminus \vec{B}},\vec{w})
}{
p(y^* | \vec{b}, \vec{v}^*_{\vec{D} \setminus \vec{B}},\vec{w}) p(z^* | \vec{b}, \vec{v}^*_{\vec{D} \setminus \vec{B}},\vec{w})
} \text{ (the first top and bottom terms cancel)}\\
&=
\log \frac{
p(\vec{b},y^*,z,\vec{v}^*_{\vec{D} \setminus \vec{B}}|\vec{w})
}{
p(\vec{b},y^*,z^*,\vec{v}^*_{\vec{D} \setminus \vec{B}}|\vec{w})
} \text{ (by the chain rule of probability)}\\
&=
H_{\vec{B} \cup \{ Z \}}(\vec{v}_{\vec{B} \cup \{ Z \}},\vec{w}) - H_{\vec{B}}(\vec{v}_{\vec{B}},\vec{w}) \text{ (by definition)}.
\end{align*}
}
Thus all terms in the curly brackets in (\ref{eqn:pairwise-cancel}) add to zero and henceforth the entire sum is zero whenever
$\vec{A}$ is not a clique.

Next, fix a term $\tilde{\phi}_{\vec{A}}(\vec{v}_{\vec{A}}, \vec{w})$ such that $\vec{A} \in {\cal C}({\cal G}_{\vec{V}})$, and fix $Y \in \vec{A}$ and $Z \in \vec{W}$ such that $Y$ and $Z$ are not adjacent in ${\cal G}(\vec{V},\vec{W})$.
We can express this term as follows:
{\small
\begin{align*}
\tilde{\phi}_{\vec{A}}(\vec{v}_{\vec{A}},\vec{w})
&=
\sum_{\vec{B} \subseteq \vec{A}} (-1)^{|\vec{A}\setminus\vec{B}|} H_{\vec{B}}(\vec{v}_{\vec{B}},\vec{w})\\
&=
\sum_{\vec{B} \subseteq \vec{A} \setminus \{ Y \}} (-1)^{|(\vec{A}\setminus\{Y\})\setminus\vec{B}|}
\left\{
H_{\vec{B}}(\vec{v}_{\vec{B}},\vec{w}) - H_{\vec{B}\cup\{Y\}}(\vec{v}_{\vec{B}\cup\{Y\}},\vec{w})
\right\}.
\end{align*}
}
We have
{\small
\begin{align*}
H_{\vec{B}}(\vec{v}_{\vec{B}},\vec{w}) - H_{\vec{B}\cup\{Y\}}(\vec{v}_{\vec{B}\cup\{Y\}},\vec{w})
&=
\log \frac{
p(y^*, \vec{v}_{\vec{B}}, \vec{v}^*_{\vec{V} \setminus (\vec{B} \cup \{ Y \})} | \vec{w})
}{
p(y, \vec{v}_{\vec{B}}, \vec{v}^*_{\vec{V} \setminus (\vec{B} \cup \{ Y \})} | \vec{w})
}
\end{align*}
}
Since $Y$ and $Z$ are not adjacent in ${\cal G}(\vec{V},\vec{W})$, by the pairwise Markov property, this object is only a function of 
$y,y^*,\vec{v}_{\vec{B}}$ and $\vec{w}_{\vec{W} \setminus \{ Z \}}$.  Applying this argument to every $Z \in \vec{W}$ that is not adjacent to some $Y \in \vec{V}$ yields that
$\tilde{\phi}_{\vec{A}}(\vec{v}_{\vec{A}},\vec{w})$ is only a function $\vec{v}_{\vec{A}}$ and $\vec{w}_{\bigcap_{C \in \vec{C}} \pa_{\cal G}(C)}$.

This establishes the result.
\end{prf}

\section{ADDITIONAL EXAMPLES OF THE L-C FACTORIZATION}
\label{supp:examples}

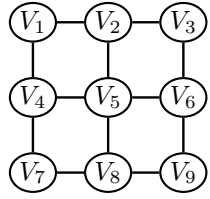
\begin{figure*}
	\begin{center}
		\begin{tikzpicture}[>=stealth, node distance=1.0cm]
		\tikzstyle{format} = [draw, thick, circle, minimum size=4.0mm,
		inner sep=1pt]
		\tikzstyle{unode} = [draw, thick, circle, minimum size=1.0mm,
		inner sep=0pt,outer sep=0.9pt]
		\tikzstyle{square} = [draw, very thick, rectangle, minimum size=4mm]

	\begin{scope}[xshift=13.7cm, node distance=1.0cm]]
		\path[-,  line width=0.9pt]
		node[format, shape=ellipse] (v1) {$V_1$}
		node[format, shape=ellipse, right of=v1, yshift=-0.0cm] (v2) {$V_2$}			
		node[format, shape=ellipse, right of=v2] (v3) {$V_3$}

		node[format, shape=ellipse, below of=v1, yshift=+0.0cm] (v4) {$V_4$}
		node[format, shape=ellipse, right of=v4, xshift=0.0cm] (v5) {$V_5$}
		node[format, shape=ellipse, right of=v5, xshift=0.0cm] (v6) {$V_6$}

		node[format, shape=ellipse, below of=v4, yshift=+0.0cm] (v7) {$V_7$}
		node[format, shape=ellipse, right of=v7, xshift=0.0cm] (v8) {$V_8$}
		node[format, shape=ellipse, right of=v8, xshift=0.0cm] (v9) {$V_9$}

		(v1) edge (v2)
		(v2) edge (v3)

		(v4) edge (v5)
		(v5) edge (v6)

		(v7) edge (v8)
		(v8) edge (v9)

		(v1) edge (v4)
		(v2) edge (v5)
		(v3) edge (v6)

		(v7) edge (v4)
		(v8) edge (v5)
		(v9) edge (v6)

		;
	\end{scope}
		\end{tikzpicture}
	\end{center}
	\caption{
	A lattice graph representing an MRF that arises in analysis of images composed of pixels, or in spin state modeling.
	}
	\label{fig:complex-graph-1}
\end{figure*}

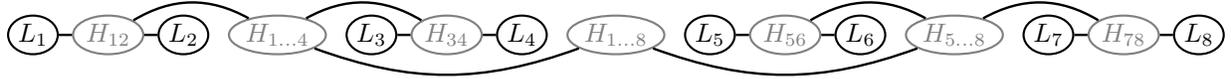
\begin{figure*}
	\begin{center}
		\begin{tikzpicture}[>=stealth, node distance=1.0cm]
		\tikzstyle{format} = [draw, thick, circle, minimum size=4.0mm,
		inner sep=1pt]
		\tikzstyle{unode} = [draw, thick, circle, minimum size=1.0mm,
		inner sep=0pt,outer sep=0.9pt]
		\tikzstyle{square} = [draw, very thick, rectangle, minimum size=4mm]

	\begin{scope}[xshift=0cm]

		\path[-,  line width=0.9pt]

			node[format, shape=ellipse, yshift=-0.0cm] (l1) {$L_1$}
			node[format, gray, shape=ellipse, right of=l1, yshift=-0.0cm] (h1) {$H_{12}$}
			node[format, shape=ellipse, right of=h1, yshift=-0.0cm] (l2) {$L_2$}

			(l1) edge (h1)
			(l2) edge (h1)

			node[format, gray, shape=ellipse, right of=l2, xshift=0.25cm, yshift=-0.0cm] (h13) {$H_{1\ldots4}$}

			node[format, shape=ellipse, right of=h13, xshift=0.25cm, yshift=-0.0cm] (l3) {$L_3$}
			node[format, gray, shape=ellipse, right of=l3, yshift=-0.0cm] (h3) {$H_{34}$}
			node[format, shape=ellipse, right of=h3, yshift=-0.0cm] (l4) {$L_4$}

			(h1) edge[bend left] (h13)
			(h3) edge[bend right] (h13)

			(l3) edge (h3)
			(l4) edge (h3)

			
			node[format, gray, shape=ellipse, right of=l4, xshift=0.25cm, yshift=-0.0cm] (h19) {$H_{1\ldots8}$}			
			

			node[format, shape=ellipse, xshift=0.25cm, right of=h19, yshift=-0.0cm] (l6) {$L_5$}
			node[format, gray, shape=ellipse, right of=l6, yshift=-0.0cm] (h67) {$H_{56}$}
			node[format, shape=ellipse, right of=h67, yshift=-0.0cm] (l7) {$L_6$}

			(l6) edge (h67)
			(l7) edge (h67)

			node[format, gray, shape=ellipse, right of=l7, xshift=0.25cm, yshift=-0.0cm] (h69) {$H_{5\ldots8}$}

			node[format, shape=ellipse, right of=h69, xshift=0.25cm, yshift=-0.0cm] (l8) {$L_7$}
			node[format, gray, shape=ellipse, right of=l8, yshift=-0.0cm] (h89) {$H_{78}$}
			node[format, shape=ellipse, right of=h89, yshift=-0.0cm] (l9) {$L_8$}

			(h67) edge[bend left] (h69)
			(h89) edge[bend right] (h69)

			(l8) edge (h89)
			(l9) edge (h89)
			
			(h13) edge[bend right=20] (h19)
			(h69) edge[bend left=20] (h19)

		;

	\end{scope}
	
		\end{tikzpicture}
	\end{center}
	\caption{
	A binary tree graph with $8$ leaf vertices corresponding to observed variables, and $7$ vertices corresponding to latent variables.
	}
	\label{fig:complex-graph-2}
\end{figure*}

The MRF model in Fig.~\ref{fig:examples} (a) generalizes to MRFs associated with lattice graphs, in particular graphs that can be represented as square grids.  Such graphs have been used to analyze particle spins
\citep{nishimori01statistical}, and in analysis of images composed of pixels \citep{szeliski22computer}.  An example of a square grid MRF model is shown in Fig.~\ref{fig:complex-graph-1}. 
The L-C factorization of this MRF model is as follows:
\begin{align*}
Z^{-1} p(v_1 \mid v_2^*, v_4^*)
p(v_2 \mid v_1^*, v_3^*, v_5^*)
p(v_3 \mid v_2^*, v_6^*)
p(v_4 \mid v_1^*, v_5^*, v_7^*)
p(v_5 \mid v_4^*, v_2^*, v_6^*, v_8^*)
p(v_6 \mid v_5^*, v_3^*, v_9^*)
p(v_7 \mid v_4^*, v_8^*) \times\\
p(v_8 \mid v_7^*, v_5^*, v_9^*)
p(v_9 \mid v_8^*, v_6^*)
OR(v_1,v_2 | v_3^*,v_4^*,v_5^*)
OR(v_2,v_3 | v_1^*,v_5^*,v_6^*)
OR(v_1,v_4 | v_2^*,v_5^*,v_7^*)
OR(v_3,v_6 | v_2^*,v_5^*,v_9^*) \times\\
OR(v_2,v_5 | v_1^*,v_4^*,v_3^*,v_6^*,v_8^*)
OR(v_4,v_5 | v_1^*,v_2^*,v_6^*,v_7^*,v_8^*)
OR(v_5,v_6 | v_2^*,v_3^*,v_4^*,v_8^*,v_9^*)
OR(v_4,v_7 | v_1^*,v_5^*,v_8^*) \times\\
OR(v_5,v_8 | v_2^*,v_4^*,v_6^*,v_7^*,v_9^*)
OR(v_6,v_9 | v_3^*,v_5^*,v_8^*)
OR(v_7, v_8 | v_4^*,v_5^*,v_9^*)
OR(v_8,v_9 | v_5^*,v_6^*,v_7^*).
\end{align*}
Since only cliques of size $1$ or $2$ exist in this graph, the factorization only contains the normalizing constant, the odds ratio terms, and the univariate conditional distribution terms.  Some reference values have been removed from these terms, without loss of generality, due to conditional independence restrictions implied by the global Markov property of this model.

Latent variable BN and MRF models have been used in the analysis of speech signals \citep{bilmes05graphical,smith11linguistic}, and philogeny \citep{durbin02biological}.  A graph representing a philogenetic tree model, where only leaf nodes
$L_1, \ldots, L_8$ are observed is shown in Fig.~\ref{fig:complex-graph-2}.
The L-C factorization of this MRF model is as follows:
\begin{align*}
Z^{-1}
p(l_1 | h_{12}^*)
p(l_2 | h_{12}^*)
p(l_3 | h_{34}^*)
p(l_4 | h_{34}^*)
p(l_5 | h_{56}^*)
p(l_6 | h_{56}^*)
p(l_7 | h_{78}^*)
p(l_8 | h_{78}^*)
p(h_{12} | l_1^*,l_2^*,h_{1\ldots4}^*)
p(h_{34} | l_3^*,l_4^*,h_{1\ldots4}^*) \times\\
p(h_{56} | l_5^*,l_6^*,h_{5\ldots8}^*)
p(h_{78} | l_7^*,l_8^*,h_{5\ldots8}^*)
p(h_{1\ldots4} | h_{12}^*,h_{34}^*,h_{1\ldots8}^*)
p(h_{5\ldots8} | h_{56}^*,h_{78}^*,h_{1\ldots8}^*)
p(h_{1\ldots8} | h_{1\ldots4}, h_{5\ldots8}) \times\\
OR(l_1,h_{12} | l_2^*,h_{1\ldots4}^*)
OR(l_2,h_{12} | l_1^*,h_{1\ldots4}^*)
OR(l_3,h_{34} | l_4^*,h_{1\ldots4}^*)
OR(l_4,h_{34} | l_3^*,h_{1\ldots4}^*)
OR(l_5,h_{56} | l_6^*,h_{5\ldots8}^*) \times\\
OR(l_6,h_{56} | l_5^*,h_{5\ldots8}^*)
OR(l_7,h_{78} | l_8^*,h_{5\ldots8}^*)
OR(l_8,h_{78} | l_7^*,h_{5\ldots8}^*)
OR(h_{12},h_{1\ldots4} | l_1^*,l_2^*,h_{34}^*,h_{1\ldots8}^*) \times\\
OR(h_{34},h_{1\ldots4} | l_3^*,l_4^*,h_{12}^*,h_{1\ldots8}^*)
OR(h_{56},h_{5\ldots8} | l_5^*,l_6^*,h_{78}^*,h_{1\ldots8}^*)
OR(h_{78},h_{5\ldots8} | l_7^*,l_8^*,h_{56}^*,h_{1\ldots8}^*) \times\\
OR(h_{1\ldots4},h_{1\ldots8} | h_{12}^*,h_{34}^*,h_{5\ldots8}^*)
OR(h_{5\ldots8},h_{1\ldots8} | h_{56}^*,h_{78}^*,h_{1\ldots4}^*).
\end{align*}
As before, since only cliques of size $1$ or $2$ exist in this graph, the factorization only contains the normalizing constant, the odds ratio terms, and the univariate conditional distribution terms.  Some reference values have been removed from these terms without loss of generality due to the global Markov property.
Prior work on philogenetic analysis has used procedures based on the structural EM algorithm to select among models of this type \citep{friedman02structural}.  An interesting area of future work is extending such methods to more general likelihoods proposed here.

\vfill

\end{document}